\documentclass[letterpaper,journal]{IEEEtran}
\ifdefined\pdfminorversion
\fi
\usepackage{amsmath,amsfonts}
\usepackage{algorithmic}
\usepackage{algorithm}
\usepackage{array}
\usepackage{multirow}
\usepackage{booktabs}
\usepackage[caption=false,font=footnotesize,labelfont=sf,textfont=sf]{subfig}
\usepackage{textcomp}
\usepackage{stfloats}
\usepackage{url}
\usepackage{verbatim}
\usepackage{graphicx}
\usepackage{xcolor}
\usepackage{cite}
\definecolor{gaincolor}{RGB}{0,128,0}
\definecolor{dropcolor}{RGB}{190,0,0}
\newcommand{\metricgain}[1]{\textcolor{gaincolor}{#1}}
\newcommand{\metricbiggain}[1]{\textcolor{gaincolor}{#1(*)}}
\newcommand{\metricsame}[1]{#1}
\newcommand{\metricdrop}[1]{\textcolor{dropcolor}{#1}}

\begin{document}

\title{CANMentor: A Large-Small Model Collaboration Framework for CAN Bus Intrusion Detection}

\author{%
Zhiquan~Liu,
Haolin~Gao,
Bo~Wu,
Jingjing~Guo,
Heng~Sun,
Jianfeng~Ma%
\thanks{This work has been submitted to the IEEE for possible publication. Copyright may be transferred without notice, after which this version may no longer be accessible.}%
\thanks{This work was supported in part by the National Natural
Science Foundation of China under Grant 62272195.}%
\thanks{Zhiquan Liu, Haolin Gao, and Bo Wu are with the College of Cyber Security,
Jinan University, Guangzhou 510632, China
(e-mail: zqliu@vip.qq.com; esiokg123gao@gmail.com; rdgalaxy520@gmail.com).}%
\thanks{Jingjing Guo and Jianfeng Ma are with the School of Cyber Engineering,
Xidian University, Xi'an 710071, China
(e-mail: jjguo@xidian.edu.cn; jfma@mail.xidian.edu.cn).}%
\thanks{Heng Sun is with the College of Information Science and Technology,
Jinan University, Guangzhou 510632, China
(e-mail: tsunheng@jnu.edu.cn).}%
} 

\markboth{Preprint}%
{Preprint}



\maketitle

\begin{abstract}
The Controller Area Network (CAN) bus lacks built-in security mechanisms, making real-time intrusion detection necessary. However, resource and latency constraints limit vehicle-side intrusion detection systems (IDSs) to lightweight models, whose detection performance and generalization can degrade under non-independent and identically distributed (non-IID) data.
Meanwhile, uploading original CAN messages to the cloud for more powerful model inference risks exposing sensitive vehicle information. We propose CANMentor, a large-small model collaboration framework in which a cloud-side large language model (LLM) teacher supervises vehicle-side lightweight IDS models for CAN bus intrusion detection. Each vehicle first performs local inference, and a privacy-aware generator then produces proxies from the original CAN messages for cloud-side teacher inference to protect data privacy. The vehicle uses the teacher prediction only when the response arrives within a prescribed latency bound and otherwise retains the local prediction; all returned teacher logits are converted into soft targets for periodic knowledge distillation. On non-IID partitions of the ROAD dataset and the HCRL Car-Hacking dataset, knowledge distillation yields maximum relative Macro F1 improvements of 25.01\% and 20.79\%, respectively. Teacher predictions also improve detection accuracy for CAN messages
near the local decision boundary by 38.71--42.00 percentage points.
\end{abstract}

\begin{IEEEkeywords}
Controller Area Network (CAN), intrusion detection systems (IDSs), non-IID, knowledge distillation, vehicular cybersecurity.
\end{IEEEkeywords}

\section{Introduction}

\IEEEPARstart{M}{odern} vehicles rely on electronic control units (ECUs) connected by in-vehicle networks (IVNs) to perform safety-critical functions. The Controller Area Network (CAN) bus remains widely used in the Internet of Vehicles (IoV) because of its low cost, reliability, and real-time performance~\cite{liuVehicularIntrusionDetection2025}. CAN does not authenticate the sender of a message. A malicious node can therefore inject forged CAN messages that receiving ECUs may act on, or flood the bus and disrupt the transmission of legitimate CAN messages~\cite{choVidenAttackerIdentification2017,tanksaleIntrusionDetectionController2024,songIntrusionDetectionSystem2016}. Detecting these intrusions on the vehicle is important because the affected functions must respond in real time.

Prior research has improved what vehicle-side intrusion detection systems (IDSs) can recognize. Learned detectors capture patterns across CAN messages~\cite{songInvehicleNetworkIntrusion2020,aungCANDIDSCANCANFD2025,aljabriLightweightAdaptiveDataDriven2025,xuMULSAMMultidimensionalAttention2025}, while message, signal, voltage, and time-frequency features provide complementary evidence of an intrusion~\cite{jeongInVehicleNetworkIntrusion2024,kangCANivalMultimodalApproach2024a,shahriarCANShieldDeepLearningBasedIntrusion2023a,dengIdentifierIDSPracticalVoltage2024,dengVSSIDSNovelIntrusion2025,liuETFIDSEntropyDrivenTimeFrequency2025}. Federated training, transfer learning, and cloud--vehicle collaboration draw on knowledge beyond a single vehicle~\cite{zhangFederatedTwostageTransformerbased2025a,abdelbassetFederatedIntrusionDetection2022,chenFedGANIDFederatedLearningBasedIntrusion2025a,zhangHybridTransferSelfSupervised2025,qinCVMIDSCloudVehicle2024}. Language models have also been adapted to detect anomalies and classify attacks in CAN sequences~\cite{nwaforCANBERTLanguagebasedIntrusion2022,alkhatibCANBERTItController2022a,liSecureBERTLlama22025}.

The difficulty arises when a vehicle encounters an attack that was rare in its own training data. Attack exposure differs across vehicles, so a lightweight IDS trained on data that are not independent and identically distributed (non-IID) can develop detection blind spots for underrepresented attack classes. A more capable cloud model could help with such decisions, but sending the original CAN messages would expose vehicle data, and waiting for a cloud response could delay the decision. The challenge is to obtain useful supervision for the current input while keeping the original CAN messages local, and to use that supervision to improve subsequent decisions on the vehicle.

\begin{figure*}[!t]
\centering
\scalebox{1}[1]{\includegraphics[width=\textwidth]{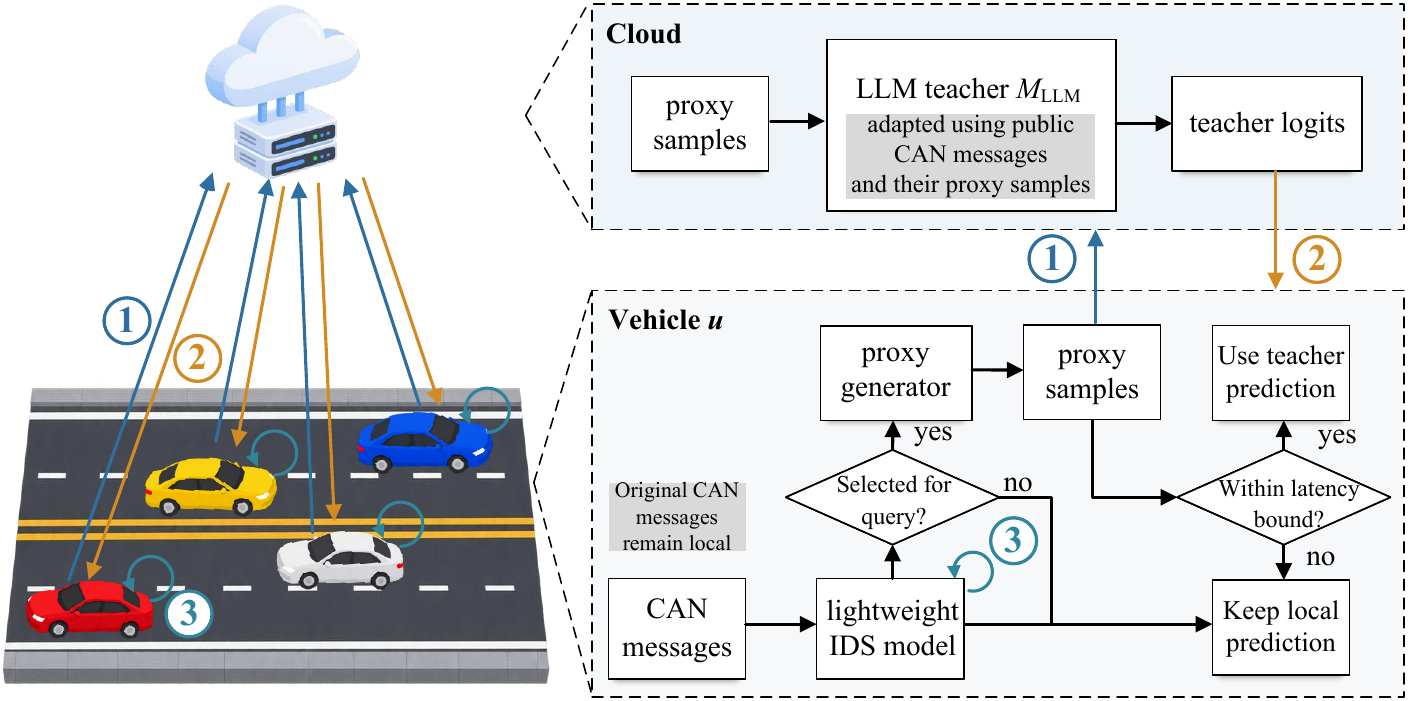}}
\caption{Overall framework of CANMentor. The numbered flows denote: 1) upload of proxy samples generated from selected CAN messages; 2) return of teacher logits and result selection; and 3) vehicle-side knowledge distillation and lightweight IDS model update. The local prediction is retained when the teacher response time exceeds the latency bound, while all returned logits are used for distillation.}
\label{fig:c2l_framework}
\end{figure*}

We propose CANMentor, a large-small model collaboration framework
for CAN bus intrusion detection, in which a cloud-side large
language model (LLM) serves as a teacher for lightweight
vehicle-side IDS models (Fig.~\ref{fig:c2l_framework}). Each vehicle first makes a local prediction. For selected inputs, it generates a proxy sample that preserves information useful for classification and sends the proxy, not the original CAN messages, to the teacher. The vehicle uses the teacher prediction if the response arrives within the prescribed latency bound; otherwise, it retains the local prediction. All returned teacher logits are converted into soft targets
for periodic knowledge distillation of the vehicle-side IDS model.

The main contributions of this paper are summarized as follows.

\begin{enumerate}
\item We propose CANMentor, a large-small model collaboration framework for CAN bus intrusion detection that combines immediate vehicle-side inference with cloud-side teacher predictions and periodic knowledge distillation from teacher soft targets.

\item We design a privacy-aware proxy generation mechanism that supports effective knowledge distillation and accurate cloud-side detection while reducing the risk of data leakage. The maximum reconstruction advantage over the no-proxy prior
is 4.22\% on ROAD at $d=0.40$ and 5.47\% on HCRL Car-Hacking
at $d=0.35$.

\item We evaluate CANMentor with three cloud-side LLM teachers on non-IID partitions of the ROAD dataset and the HCRL Car-Hacking dataset. Knowledge distillation yields maximum relative Macro F1 improvements of 25.01\% and 20.79\% on the two datasets, respectively. Teacher predictions on proxy samples corresponding to
low-confidence inputs outperform the local predictions
by 38.71--42.00 percentage points.

\end{enumerate}

\section{Related Work}

\subsection{Vehicle-Side CAN Bus Intrusion Detection}
Vehicle-side IDSs can identify an intrusion from the sender's physical characteristics or from changes in traffic behavior. Viden uses transmitter voltages to identify attacker ECUs~\cite{choVidenAttackerIdentification2017}. Taylor et al. and Song et al. examine message frequencies and inter-arrival times, respectively~\cite{taylorFrequencybasedAnomalyDetection2015,songIntrusionDetectionSystem2016}.

Learned detectors use CAN traffic itself to identify attack patterns, including through deep convolutional networks and frame-aware features~\cite{songInvehicleNetworkIntrusion2020,amatoCANBusAttackDetection2021,jeongInVehicleNetworkIntrusion2024,tanksaleIntrusionDetectionController2024}. Reviews discuss automotive IDS design and explainability in connected vehicles~\cite{lampeIntrusionDetectionAutomotive2023,nwakanmaExplainableArtificialIntelligence2023}. Other work addresses deployment through lightweight detection, hardware acceleration, or support for both CAN and CAN with flexible data rate (CAN-FD)~\cite{aljabriLightweightAdaptiveDataDriven2025,xuMULSAMMultidimensionalAttention2025,aungCANDIDSCANCANFD2025}. Efficient local inference addresses the cost of executing the detector. A separate problem is how that detector learns to recognize attacks scarcely represented in its local training set.

\subsection{Collaborative IDS Under Non-IID Data}

Collaborative IDS methods leverage knowledge across vehicles or cloud-side components to mitigate the limitations of isolated local training under non-IID data. Abdel-Basset et al. applied federated intrusion detection to smart transportation systems~\cite{abdelbassetFederatedIntrusionDetection2022}, while Zhang et al. developed a federated two-stage Transformer for non-IID CAN bus intrusion detection~\cite{zhangFederatedTwostageTransformerbased2025a}. CVMIDS maps CAN messages from different vehicles into a shared multidimensional feature space for cloud-side model training~\cite{qinCVMIDSCloudVehicle2024}. FedGAN-ID combines federated learning with differentially
private synthetic attack data generated by a generative
adversarial network (GAN) to address the scarcity of attack
samples and mitigate the effects of non-IID data across
vehicles~\cite{chenFedGANIDFederatedLearningBasedIntrusion2025a}. These studies primarily improve collaborative training through federated optimization, shared representations, or privacy-preserving data augmentation. Taken together, existing studies mainly focus on either privacy-preserving
collaborative training or cloud-assisted detection. At their intersection, CANMentor introduces a privacy-preserving
collaboration mechanism that uses proxy samples within a
cloud--vehicle teacher--student framework for online detection
and continued vehicle-side model improvement under latency constraints.

\subsection{Language-Model-Based CAN Bus Intrusion Detection}

Transformer-based language models capture long-range dependencies in CAN sequences through self-attention~\cite{vaswaniAttentionAllYou2017}, while low-rank adaptation (LoRA) reduces the cost of downstream adaptation~\cite{huLoRALowRankAdaptation}.
Nwafor and Olufowobi proposed CANBERT for supervised intrusion classification
using CAN message fields, whereas Alkhatib et al. proposed CAN-BERT for
self-supervised anomaly detection over CAN identifier
sequences~\cite{nwaforCANBERTLanguagebasedIntrusion2022,alkhatibCANBERTItController2022a}.

Subsequent systems include CANLP, which combines natural language processing
with embedded machine learning~\cite{balasubramanianCANLPIntrusionDetection2025};
IoV-BERT-IDS, a hybrid LLM-based IDS for IoV
environments~\cite{fuIoVBERTIDSHybridNetwork2025}; and CAN-C-BERT,
CAN-SecureBERT, and CAN-LLAMA2, which adapt pretrained language models to CAN
bus intrusion detection~\cite{liSecureBERTLlama22025}. A recent survey further identifies generative AI as an emerging direction for
IVN intrusion detection~\cite{asaouerGenerativeAIBased2026}. CANMentor focuses
on a cloud-side LLM teacher that supports vehicle-side detection and periodic
knowledge distillation.

\section{Preliminaries}
\label{sec:preliminaries}

\subsection{CAN Frame Format}

\begin{figure}[!t]
\centering
\includegraphics[width=\columnwidth]{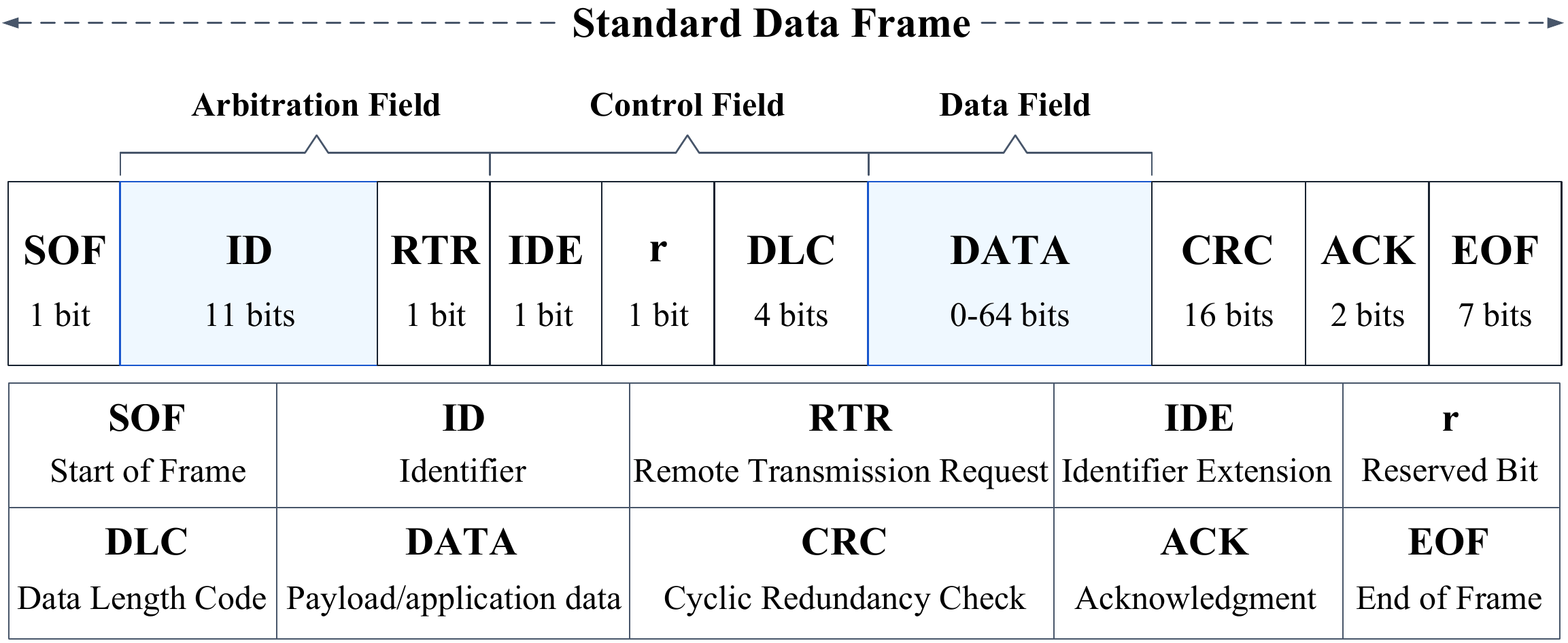}
\caption{CAN frame format.}
\label{fig:fig1}
\end{figure}

CAN uses broadcast communication, allowing all ECUs on a bus to receive each transmitted frame. As shown in Fig.~\ref{fig:fig1}, a standard Classical CAN data frame contains framing, arbitration, control, data, cyclic redundancy check (CRC), and acknowledgment fields~\cite{liuVehicularIntrusionDetection2025}. Its 11-bit identifier determines message type and arbitration priority, while the data field carries up to 8 bytes. Identifiers, payloads, timestamps, and inter-arrival times provide content and temporal features for intrusion detection. Because individual frames provide limited context, sequence analysis helps identify anomalous communication patterns.

\subsection{Adversary Model}
\label{sec:adversary_model}

We consider an adversary that has gained access to the in-vehicle CAN bus
and can observe traffic, inject malicious CAN messages, and, for masquerade
attacks, suppress and replace legitimate transmissions. The adversary may
manipulate CAN identifiers, payloads, or transmission patterns~\cite{wooPracticalWirelessAttack2015,sunClassificationAttacks2020}. We consider
the following attack capabilities.
\subsubsection{Flooding and Fuzzing Attacks}
The adversary can inject high-priority messages at a high rate to disrupt
legitimate communication, as in denial-of-service (DoS) attacks, or inject
randomly constructed messages with arbitrary identifiers and payloads, as in
fuzzing attacks.
\subsubsection{Spoofing Attacks}
The adversary can identify a target CAN identifier and inject fabricated
messages with manipulated payloads to alter the corresponding vehicle state,
while legitimate messages associated with the same identifier may remain on
the bus.
\subsubsection{Masquerade Attacks}
The adversary can suppress legitimate messages associated with a target CAN
identifier and replace them with manipulated messages while preserving the
expected transmission pattern, thereby altering vehicle-state information and
misleading receiving ECUs.

\subsection{Wasserstein GAN With Gradient Penalty}
\label{sec:wgan_gp}
Wasserstein GAN (WGAN) optimizes the Wasserstein-1 distance through a real-valued critic to provide informative gradients and stabilize training. Wasserstein GAN with gradient penalty (WGAN-GP) replaces weight clipping with a gradient penalty that encourages a 1-Lipschitz critic by keeping its gradient norm close to 1~\cite{gulrajaniImprovedTrainingWasserstein2017}. A generator $G$ maps noise $z$ to generated samples $\widetilde{x}=G(z)$, and a critic $D$ assigns higher scores to real samples than to generated samples. The critic loss is
\begin{equation}
\label{eq:wgan_gp_critic}
\begin{aligned}
\mathcal{L}_{\mathrm{D}}
&=
\mathbb{E}_{\widetilde{x}\sim p_g}[D(\widetilde{x})]
-
\mathbb{E}_{x\sim p_r}[D(x)] \\
&\quad+
\lambda
\mathbb{E}_{\hat{x}\sim p_{\hat{x}}}
\left[
\left(
\left\|
\nabla_{\hat{x}}D(\hat{x})
\right\|_2
-
1
\right)^2
\right],
\end{aligned}
\end{equation}
where $p_r$ and $p_g$ are the real and generated data distributions, respectively, $\lambda$ is the gradient penalty coefficient, and $\hat{x}\sim p_{\hat{x}}$ is sampled along straight lines between real and generated samples. CANMentor uses WGAN-GP to train the vehicle-side proxy generator.

\subsection{Large Language Models and LoRA Fine-Tuning}

Large language models (LLMs) are typically Transformer-based models trained on large-scale text corpora. Their sequence modeling capability enables them to learn contextual representations and capture dependencies within sequences~\cite{vaswaniAttentionAllYou2017,devlinBERTPretrainingDeep2019}. Low-rank adaptation (LoRA) is a parameter-efficient method that adapts a pretrained LLM by freezing its weights and learning low-rank update matrices for downstream tasks~\cite{huLoRALowRankAdaptation}. For a frozen weight matrix $\mathbf{W}_0\in\mathbb{R}^{d\times k}$, the update is
\begin{equation}
\mathbf{W}
=
\mathbf{W}_0+\Delta\mathbf{W},
\quad
\Delta\mathbf{W}
=
\frac{\alpha}{r}\mathbf{B}\mathbf{A},
\end{equation}
where $\mathbf{B}\in\mathbb{R}^{d\times r}$, $\mathbf{A}\in\mathbb{R}^{r\times k}$, $r\ll\min(d,k)$ is the LoRA rank, and $\alpha$ is a scaling factor. During fine-tuning, $\mathbf{W}_0$ remains frozen, while $\mathbf{A}$ and $\mathbf{B}$ are updated.

\subsection{Knowledge Distillation}

Knowledge distillation transfers predictive knowledge from a teacher to a student suited to deployment using soft targets that convey the teacher's relative class confidence~\cite{hintonDistillingKnowledgeNeural2015}. Temperature-scaled softmax with temperature $\tau$ converts teacher logits $\boldsymbol{\ell}^{\mathrm{T}}$ and student logits $\boldsymbol{\ell}^{\mathrm{S}}$ into soft targets $\mathbf{p}^{\mathrm{T}}_{\tau}$ and student probabilities $\mathbf{p}^{\mathrm{S}}_{\tau}$, respectively, over $K$ classes; larger $\tau$ produces smoother targets. Transfer minimizes the Kullback--Leibler divergence $D_{\mathrm{KL}}(\mathbf{p}^{\mathrm{T}}_{\tau}\parallel\mathbf{p}^{\mathrm{S}}_{\tau})$ from teacher soft targets to student probabilities.

\section{Method}
The main notation used in this section is summarized in Table~\ref{tab:notation}.
\begin{table}[!t]
\centering
\caption{Main notation used in CANMentor}
\label{tab:notation}
\fontsize{7}{8.4}\selectfont
\setlength{\tabcolsep}{2pt}
\renewcommand{\arraystretch}{1.03}
\begin{tabular}{@{}
  >{\raggedright\arraybackslash}p{0.16\columnwidth}
  >{\raggedright\arraybackslash}p{\dimexpr0.34\columnwidth-6pt\relax}
  >{\raggedright\arraybackslash}p{0.18\columnwidth}
  >{\raggedright\arraybackslash}p{\dimexpr0.32\columnwidth-6pt\relax}@{}}
\toprule
Symbol & Description & Symbol & Description \\
\midrule
$L,\ K$
& Window length; number of classes
& $s_i^{(u)},\ \theta_{\mathrm{conf}}$
& Local prediction confidence; confidence threshold \\
$\mathcal{D}_u,\ \mathcal{D}_{\mathrm{pub}}$
& Local dataset of windowed samples; labeled public CAN dataset
& $d,\ d_i^{(u)}$
& Prescribed perturbation distance; mean absolute perturbation distance \\
$\mathbf{X}_i^{(u)},\ \widetilde{\mathbf{X}}_i^{(u)}$
& Original windowed sample; corresponding proxy sample
& $\boldsymbol{\ell}^{\mathrm{S}},\ \boldsymbol{\ell}^{\mathrm{T}},\ \boldsymbol{\ell}^{\mathrm{A}}$
& Local detection logits; teacher logits; anchor logits \\
$C_u,\ C_u^{0}$
& Lightweight IDS model; its frozen copy before distillation
& $t_i^{(u)},\ T_{\max}$
& Response time; prescribed latency bound \\
$G_u,\ D_u$
& Proxy generator; WGAN-GP critic
& $\tau$
& Distillation temperature \\
$M_{\mathrm{LLM}}$
& Complete teacher mapping
& $H_i^{\mathrm{T},(u)},\ \omega_i^{(u)}$
& Entropy of the\newline teacher prediction;\newline corresponding~weight \\
\bottomrule
\end{tabular}
\end{table}

\subsection{System Model}

As shown in Fig.~\ref{fig:c2l_framework}, CANMentor comprises vehicles equipped with lightweight IDS models and proxy generators, and a cloud-side LLM teacher. Each vehicle first performs local inference and generates proxy samples from selected CAN messages, preserving class-relevant behavior while retaining raw messages locally.

Before online collaboration, the teacher is fine-tuned and adapted in a trusted offline environment, then deployed to the cloud. It returns logits for each uploaded proxy sample. The vehicle uses a teacher prediction only if it arrives within the prescribed latency bound; otherwise, it retains the local prediction. All returned logits support periodic knowledge distillation of the vehicle-side IDS models.

\subsection{Data Preprocessing}
\label{sec:data_preprocessing}

CANMentor represents each CAN message using its identifier, payload, and inter-arrival time. Let $m_i=(t_i,\mathrm{ID}_i,c_i)$ denote the $i$-th message, where
$t_i$, $\mathrm{ID}_i$, and $c_i$ are its timestamp, CAN identifier,
and payload, respectively. The standard 11-bit identifier and payload are encoded and concatenated as
\begin{equation}
\begin{aligned}
\mathbf{b}^{\mathrm{ID}}_i
&=
\operatorname{bin}_{11}(\mathrm{ID}_i) \in \{0,1\}^{11},\\
\mathbf{b}^{\mathrm{pay}}_i
&=
\operatorname{bin}_{64}(c_i) \in \{0,1\}^{64},\\
\mathbf{b}_i
&=
[\mathbf{b}^{\mathrm{ID}}_i; \mathbf{b}^{\mathrm{pay}}_i]
\in \{0,1\}^{75}.
\end{aligned}
\end{equation}

For payloads shorter than 8 bytes, valid bytes are retained according to the data length code (DLC), and zero padding maintains a fixed length. The normalized inter-arrival time is appended to each encoded message, and sliding windows group $L$ consecutive messages:
\begin{equation}
\begin{aligned}
\Delta t_i
&=
t_i - t_{i-1},\\
\widehat{\Delta t}_i
&=
\frac{\Delta t_i-\mu_{\Delta t}}
{\sigma_{\Delta t}+\epsilon},\\
\mathbf{x}_i
&=
[\mathbf{b}_i; \widehat{\Delta t}_i]
\in \mathbb{R}^{76},\\
\mathbf{X}_i
&=
[\mathbf{x}_{i-L+1}, \mathbf{x}_{i-L+2}, \ldots, \mathbf{x}_i]
\in \mathbb{R}^{L \times 76}.
\end{aligned}
\end{equation}

Here, $\mu_{\Delta t}$ and $\sigma_{\Delta t}$ are the mean
and standard deviation of inter-arrival times computed on the
training set, and $\epsilon>0$ is a small constant preventing
division by zero. For vehicle $u$, the resulting $N_u$ windowed samples form $\mathcal{D}_u=\{\mathbf{X}^{(u)}_i\}_{i=1}^{N_u}$, with $\mathbf{X}^{(u)}_i\in\mathbb{R}^{L\times76}$.

\subsection{Vehicle-Side Detection and Proxy Sample Generation}

Vehicle $u$ uses a lightweight IDS model $C_u$ and a proxy generator $G_u$.
Any IDS architecture that outputs detection logits over $K$ classes can be
used. For an input windowed sample $\mathbf{X}^{(u)}_i \in \mathcal{D}_u$,
the local detection logits, class probabilities, and predicted label are
\begin{equation}
\begin{aligned}
\boldsymbol{\ell}^{\mathrm{S},(u)}_i &= C_u(\mathbf{X}^{(u)}_i),\quad
\boldsymbol{\pi}^{\mathrm{S},(u)}_i = \operatorname{softmax}(\boldsymbol{\ell}^{\mathrm{S},(u)}_i),\\
\hat{y}^{\mathrm{S},(u)}_i &= \operatorname*{arg\,max}_{k}\boldsymbol{\pi}^{\mathrm{S},(u)}_{i,k}.
\end{aligned}
\end{equation}

CANMentor queries the cloud-side teacher for windows on which the local IDS has low confidence. Under non-IID data, attacks underrepresented in local training can also receive incorrect predictions with high confidence. To sample such cases, CANMentor randomly selects a small proportion of the remaining windows. For $\mathbf{X}^{(u)}_i \in \mathcal{D}_u$, the local prediction confidence is measured by the maximum class probability:
\begin{equation}
s^{(u)}_i
=
\max_k \boldsymbol{\pi}^{\mathrm{S},(u)}_{i,k}.
\end{equation}
A windowed sample is selected by the low-confidence criterion when
$s^{(u)}_i < \theta_{\mathrm{conf}}$; the remaining samples are eligible for
random selection.

\begin{figure}[!t]
\centering
\includegraphics[width=\columnwidth]{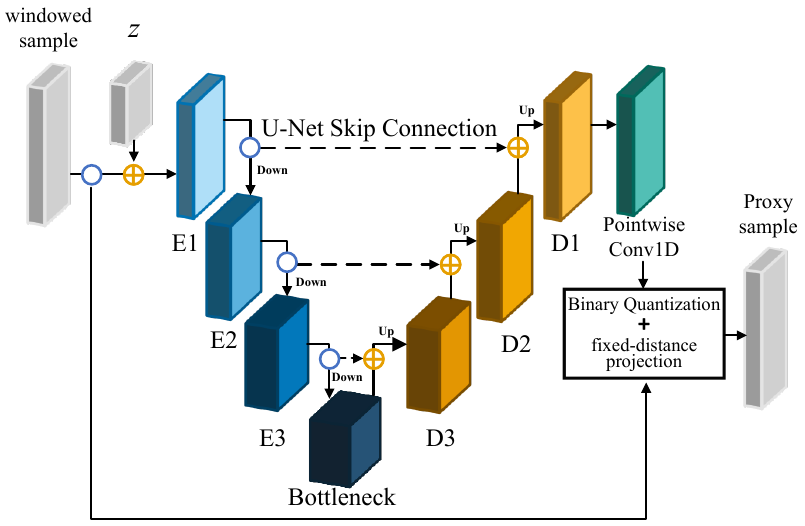}
\caption{Architecture of the noise-conditioned U-Net proxy generator. Dashed lines indicate encoder--decoder skip connections. Down and Up denote stride-2 convolution and linear upsampling, respectively. The output stage enforces the prescribed perturbation distance.}
\label{fig:generator}
\end{figure}

CANMentor uploads only proxy samples while retaining the corresponding windowed samples on the vehicle. As shown in Fig.~\ref{fig:generator}, $G_u$ consists of a noise-conditioned 1D U-Net with a three-stage encoder, a symmetric decoder, and cross-scale skip connections, followed by pointwise convolution, binary quantization, and fixed-distance projection. For a selected windowed sample $\mathbf{X}^{(u)}_i$, a noise vector
$\mathbf{z}^{(u)}_i$, and a prescribed perturbation distance $d$, the mapping is
\begin{equation}
\widetilde{\mathbf{X}}^{(u)}_i
=
G_u\left(\mathbf{X}^{(u)}_i,\mathbf{z}^{(u)}_i;d\right),
\quad
\mathbf{z}^{(u)}_i\sim\mathcal{N}(\mathbf{0},\mathbf{I}).
\end{equation}

Following the WGAN-GP formulation in Section~\ref{sec:wgan_gp},
vehicle $u$ trains $G_u$ to match the distribution of real
windowed samples. The generator
adversarial loss is
\begin{equation}
\mathcal{L}_{\mathrm{adv}}
=
-
\mathbb{E}
\left[
 D_u(\widetilde{\mathbf{X}}^{(u)}_i)
\right],
\end{equation}
where $D_u$ is the WGAN-GP critic.

To preserve class-relevant behavior, $C_u$ is frozen during proxy generator training. CANMentor introduces the semantic-consistency loss $\mathcal{L}_{\mathrm{sem}}$ and feature-consistency loss $\mathcal{L}_{\mathrm{feat}}$. These losses constrain each proxy sample to match the prediction behavior and intermediate representation of its source windowed sample:
\begin{equation}
\begin{aligned}
\mathcal{L}_{\mathrm{sem}}
&=
\tau^2
D_{\mathrm{KL}}
\Bigl(
\operatorname{softmax}
\bigl(C_u(\mathbf{X}^{(u)}_i)/\tau\bigr)
\\
&\quad\parallel
\operatorname{softmax}
\bigl(C_u(\widetilde{\mathbf{X}}^{(u)}_i)/\tau\bigr)
\Bigr),\\
\mathcal{L}_{\mathrm{feat}}
&=
\left\|
f_u(\widetilde{\mathbf{X}}^{(u)}_i)
-
f_u(\mathbf{X}^{(u)}_i)
\right\|_2^2.
\end{aligned}
\end{equation}
Here, $\tau$ is the temperature parameter, and $f_u(\cdot)$ denotes an intermediate feature layer of $C_u$.

The distance loss controls how much each proxy differs from its source window. Larger perturbations obscure more of the original bit pattern but can also disrupt features used for classification. We measure the mean absolute perturbation distance $d^{(u)}_i$ and penalize distances outside the interval $[m_{\min},m_{\max}]$:
\begin{equation}
\begin{aligned}
d^{(u)}_i
&=
\operatorname{mean}
\left(
\left|
\widetilde{\mathbf{X}}^{(u)}_i
-
\mathbf{X}^{(u)}_i
\right|
\right),\\
\mathcal{L}_{\mathrm{dist}}
&=
\max(0,m_{\min}-d^{(u)}_i)
+
\max(0,d^{(u)}_i-m_{\max}).
\end{aligned}
\end{equation}

The overall loss for training the proxy generator is
\begin{equation}
\begin{aligned}
\mathcal{L}_{\mathrm{G}}
=&
\underbrace{
\lambda_{\mathrm{adv}}\mathcal{L}_{\mathrm{adv}}
}_{\text{adversarial loss}}
+
\underbrace{
\lambda_{\mathrm{sem}}\mathcal{L}_{\mathrm{sem}}
+
\lambda_{\mathrm{feat}}\mathcal{L}_{\mathrm{feat}}
}_{\text{semantic and feature consistency losses}}
\\
&+
\underbrace{
\lambda_{\mathrm{dist}}\mathcal{L}_{\mathrm{dist}}
}_{\text{distance loss}}.
\end{aligned}
\end{equation}
$G_u$ is trained by minimizing $\mathcal{L}_{\mathrm{G}}$. During training,
$\mathcal{L}_{\mathrm{dist}}$ softly guides the perturbation distance. Because
this loss cannot guarantee the prescribed distance for every uploaded proxy
sample, the generator output stage enforces $d^{(u)}_i=d$ as a hard constraint.
Specifically, it converts $d$ into the required number of bit flips and flips
the positions with the highest generator-predicted flip probabilities.

\subsection{Cloud-Side Teacher Inference and Knowledge Distillation}
\subsubsection{Numerical CAN Frame Encoding and Teacher Fine-Tuning}

CAN messages are commonly recorded in textual form with a timestamp and
hexadecimal representations of the CAN identifier and payload. Directly
concatenating these fields for LLM tokenization is problematic because
language tokenizers are designed for natural language and may split
hexadecimal fields into variable-length units without explicitly preserving
the field boundaries or bit-level relationships within a CAN frame. Prior
studies address similar representation mismatches using learnable encoders
that map numerical time-series data into the embedding space of pretrained
language models~\cite{sunTESTTextPrototype2024,chenLargeLanguageModels2025}.

Following this idea, CANMentor introduces a numerical CAN frame
encoder $\mathcal{A}_{\eta}$, hereafter referred to as the numerical encoder,
before the cloud-side LLM teacher. For each frame $\mathbf{x}_i$, an identifier encoder processes
$\mathbf{b}^{\mathrm{ID}}_i$, a payload-byte encoder processes the eight
bytes in $\mathbf{b}^{\mathrm{pay}}_i$, and a temporal encoder processes
$\widehat{\Delta t}_i$. Learnable position embeddings distinguish the eight
payload-byte positions. The resulting field
embeddings are concatenated and projected into the hidden dimension of the
LLM. A residual projection of $\mathbf{x}_i$ is further added to retain the
complete frame-level information.

For a windowed sample $\mathbf{X}^{(u)}_i$, the numerical encoder produces one continuous embedding for each frame and appends a learnable classification token:
\begin{equation}
\mathbf{E}^{(u)}_i
=
\mathcal{A}_{\eta}
\left(
\mathbf{X}^{(u)}_i
\right)
=
\left[
\mathbf{e}^{(u)}_{i-L+1},
\ldots,
\mathbf{e}^{(u)}_i,
\mathbf{e}_{\mathrm{cls}}
\right],
\end{equation}
where $\mathbf{E}^{(u)}_i\in\mathbb{R}^{(L+1)\times d_h}$ and $d_h$ is the hidden dimension of the LLM. The resulting embeddings are supplied directly to the embedding interface of the LLM without textual serialization or tokenization.

The cloud-side LLM teacher processes the encoded sequence, and a classification head $H_{\psi}$ maps the hidden state of the classification token to the teacher logits:
\begin{equation}
\boldsymbol{\ell}^{\mathrm{T},(u)}_i
=
H_{\psi}
\left(
M_{\theta,\phi}
\left(
\mathbf{E}^{(u)}_i
\right)_{\mathrm{cls}}
\right),
\end{equation}
where $\theta$ denotes the frozen pretrained parameters of the LLM and $\phi$ denotes its trainable LoRA parameters.

The numerical encoder, LoRA parameters, and classification head are jointly fine-tuned using a large labeled public CAN dataset
$\mathcal{D}_{\mathrm{pub}}
=
\{(\mathbf{X}^{\mathrm{pub}}_j,y^{\mathrm{pub}}_j)\}_{j=1}^{N_{\mathrm{pub}}}$.
The corresponding classification objective is
\begin{equation}
\mathcal{L}_{\mathrm{base}}
=
\frac{1}{N_{\mathrm{pub}}}
\sum_{j=1}^{N_{\mathrm{pub}}}
\mathcal{L}_{\mathrm{CE}}
\left(
\boldsymbol{\ell}^{\mathrm{T}}_j,
y^{\mathrm{pub}}_j
\right).
\end{equation}
Here, $\mathcal{L}_{\mathrm{CE}}$ denotes the cross-entropy loss.
After fine-tuning, the teacher can directly capture discriminative patterns in CAN messages and provide class-level supervision.

\subsubsection{Teacher Adaptation to Proxy Samples}

For brevity, $M_{\mathrm{LLM}}$ denotes the complete teacher mapping,
including $\mathcal{A}_{\eta}$, $M_{\theta,\phi}$, and $H_{\psi}$.

The teacher is initially fine-tuned on original public windowed
samples but receives proxy samples during online inference.
Perturbations introduced by the proxy generator create a
distribution shift between these inputs. To address this shift,
the teacher is adapted using proxy samples generated from public
CAN data once per deployed generator version and reused until
that generator is updated. We assume that initial fine-tuning and all proxy adaptation occur in a trusted offline environment (e.g., a vehicle manufacturer's server). Only the resulting teacher weights are transferred to the inference cloud.

For proxy adaptation, each current generator $G_u$ transforms the
$N_{\mathrm{adapt}}$ labeled public windowed samples into proxy samples
$\widetilde{\mathbf{X}}^{\mathrm{pub},(u)}_j$ at the prescribed distance
$d$, while retaining the corresponding labels
$y^{\mathrm{pub}}_j$.
The teacher is then adapted using both the generated proxy samples and
the corresponding public windowed samples:
\begin{equation}
\begin{aligned}
\mathcal{L}_{\mathrm{adapt}}
&=
\frac{1}{|\mathcal{U}|N_{\mathrm{adapt}}}
\sum_{u\in\mathcal{U}}
\sum_{j=1}^{N_{\mathrm{adapt}}}
\\[-1mm]
&\quad\times
\Big[
\lambda_{\mathrm{proxy}}
\mathcal{L}_{\mathrm{CE}}
\big(
M_{\mathrm{LLM}}
\left(\widetilde{\mathbf{X}}^{\mathrm{pub},(u)}_j\right),
y^{\mathrm{pub}}_j
\big)
\\[-1mm]
&\qquad+
\lambda_{\mathrm{clean}}
\mathcal{L}_{\mathrm{CE}}
\big(
M_{\mathrm{LLM}}(\mathbf{X}^{\mathrm{pub}}_j),
y^{\mathrm{pub}}_j
\big)
\Big].
\end{aligned}
\end{equation}
Here, $\mathcal{U}$ denotes the set of vehicles, and $\lambda_{\mathrm{proxy}}$ and $\lambda_{\mathrm{clean}}$ are
weighting coefficients. This joint objective enables the teacher to identify class-discriminative attack patterns in proxy samples while retaining its detection capability on original windowed samples.

\subsubsection{Cloud-Side Teacher Inference and Vehicle-Side Knowledge Distillation}

For each selected windowed sample $\mathbf{X}^{(u)}_i$, vehicle $u$
uploads the corresponding proxy sample
$\widetilde{\mathbf{X}}^{(u)}_i$. The adapted cloud-side LLM teacher
produces the teacher logits
\begin{equation}
    \boldsymbol{\ell}^{\mathrm{T},(u)}_i
    =
    M_{\mathrm{LLM}}
    \left(\widetilde{\mathbf{X}}^{(u)}_i\right).
\end{equation}
Let $t_{\mathrm{veh},i}^{(u)}$ denote the vehicle-side processing time,
$t_{\mathrm{comm},i}^{(u)}$ the round-trip communication time, and
$t_{\mathrm{LLM},i}^{(u)}$ the teacher inference time.
The response time is
\begin{equation}
t_i^{(u)}
=
t_{\mathrm{veh},i}^{(u)}
+t_{\mathrm{comm},i}^{(u)}
+t_{\mathrm{LLM},i}^{(u)}.
\end{equation}
Given a prescribed latency bound $T_{\max}$, vehicle $u$ selects the current detection result as
\begin{equation}
\hat{y}^{(u)}_i
=
\begin{cases}
\operatorname*{arg\,max}_{k}
\ell^{\mathrm{T},(u)}_{i,k},
& t_i^{(u)}\leq T_{\max},\\
\hat{y}^{\mathrm{S},(u)}_i,
& t_i^{(u)}>T_{\max}.
\end{cases}
\end{equation}
Regardless of the response time, all returned teacher logits are retained for
periodic knowledge distillation. During distillation, vehicle $u$ converts
them into soft targets using temperature-scaled softmax and updates $C_u$ on
the corresponding selected windowed samples retained locally.

To account for differences in teacher prediction confidence, CANMentor
converts the teacher logits into class probabilities
$\mathbf{p}^{\mathrm{T},(u)}_i
=
\operatorname{softmax}(\boldsymbol{\ell}^{\mathrm{T},(u)}_i)$.
The entropy of the teacher prediction and its corresponding weight are
\begin{equation}
\begin{aligned}
H^{\mathrm{T},(u)}_i
&=
-\sum_{k=1}^{K}
p^{\mathrm{T},(u)}_{i,k}
\log p^{\mathrm{T},(u)}_{i,k},\\
\omega^{(u)}_i
&=
\max\left\{
0,\,
1-\frac{H^{\mathrm{T},(u)}_i}{\log K}
\right\}.
\end{aligned}
\end{equation}

Let $N^{\mathrm{sel}}_u$ denote the number of selected windowed samples
at vehicle $u$. The entropy-weighted distillation loss is
\begin{equation}
\begin{aligned}
\mathcal{L}_{\mathrm{teacher}}^{(u)}
&=
\frac{\tau^2}{\sum_i\omega^{(u)}_i+\epsilon}
\sum_i
\omega^{(u)}_i
D_{\mathrm{KL}}
\Bigl(
\operatorname{softmax}
(\boldsymbol{\ell}^{\mathrm{T},(u)}_i/\tau)
\\
&\quad\parallel
\operatorname{softmax}
(\boldsymbol{\ell}^{\mathrm{S},(u)}_i/\tau)
\Bigr),
\end{aligned}
\end{equation}
where the sums are taken over the selected windowed samples.

Directly distilling the teacher soft targets may shift the existing
detection behavior of the lightweight IDS model. CANMentor therefore
retains a frozen copy $C_u^{0}$ of the model before distillation and
computes its anchor logits as
$\boldsymbol{\ell}^{\mathrm{A},(u)}_i
=
C_u^{0}(\mathbf{X}^{(u)}_i)$.
The anchor loss and the complete distillation objective are
\begin{equation}
\begin{aligned}
\mathcal{L}_{\mathrm{anchor}}^{(u)}
&=
\frac{\tau^2}{N^{\mathrm{sel}}_u}
\sum_i
D_{\mathrm{KL}}
\Bigl(
\operatorname{softmax}
(\boldsymbol{\ell}^{\mathrm{A},(u)}_i/\tau)
\\
&\quad\parallel
\operatorname{softmax}
(\boldsymbol{\ell}^{\mathrm{S},(u)}_i/\tau)
\Bigr),\\
\mathcal{L}_{\mathrm{KD}}^{(u)}
&=
\mathcal{L}_{\mathrm{teacher}}^{(u)}
+
\lambda_{\mathrm{anchor}}
\mathcal{L}_{\mathrm{anchor}}^{(u)}.
\end{aligned}
\end{equation}
Here, $\tau$ is the distillation temperature, $\epsilon$ is a small
numerical constant, and $\lambda_{\mathrm{anchor}}$ controls the strength
of the anchor loss.

\section{Experiments}
\subsection{Datasets}

We use the ROAD and HCRL Car-Hacking datasets to evaluate CANMentor
on the attack types described in Section~\ref{sec:adversary_model}
\cite{seoGIDSGANBased2018,vermaComprehensiveGuideCAN2024}.
HCRL Car-Hacking covers flooding, fuzzing, and spoofing attacks,
whereas ROAD covers fuzzing and simulated masquerade attacks
targeting selected vehicle signals.
\subsubsection{ROAD Dataset}
\begin{table}[!t]
\centering
\caption{Statistics of the ROAD dataset}
\label{tab:road_dataset}
\fontsize{8}{9.6}\selectfont
\begin{tabular}{lc}
\hline
Attack class & \begin{tabular}[c]{@{}c@{}}Number of\\attack messages\end{tabular} \\
\hline
Fuzzing attack & 1,059 \\
Correlated signal fabrication & 5,493 \\
Correlated signal masquerade & 5,493 \\
Max engine coolant temperature fabrication & 43 \\
Max engine coolant temperature masquerade & 43 \\
Max speedometer fabrication & 11,694 \\
Max speedometer masquerade & 11,694 \\
Reverse light off fabrication & 5,480 \\
Reverse light off masquerade & 5,480 \\
Reverse light on fabrication & 8,034 \\
Reverse light on masquerade & 8,034 \\
\hline
Total & 62,547 \\
\hline
\end{tabular}
\end{table}
The Real ORNL Automotive Dynamometer (ROAD) CAN Intrusion Dataset was
collected from a real vehicle under ambient driving and physically verified
attack conditions~\cite{vermaComprehensiveGuideCAN2024}.
It contains 12 ambient captures totaling about 3 hours and 33 attack captures
totaling about 30 minutes, covering fuzzing and five targeted attack classes:
correlated signal, max engine coolant temperature, max speedometer, reverse
light off, and reverse light on.

In our evaluation, the simulated masquerade variants of the five targeted
attack classes represent the masquerade capability defined in
Section~\ref{sec:adversary_model}. Table~\ref{tab:road_dataset} summarizes
the publicly labeled attack captures. Accelerator attack captures are excluded
because their labels are not publicly available.

\subsubsection{HCRL Car-Hacking Dataset}
The HCRL Car-Hacking dataset, used in GIDS~\cite{seoGIDSGANBased2018}, was collected from a real vehicle through the on-board diagnostics II (OBD-II) port during CAN message injection attacks. It contains denial-of-service (DoS), Fuzzy, gear spoofing, and revolutions-per-minute (RPM) spoofing attacks. Each attack file includes normal and injected CAN messages; Table~\ref{tab:hcrl_dataset} summarizes the dataset statistics.

\begin{table}[!t]
\centering
\caption{Statistics of the HCRL Car-Hacking dataset}
\label{tab:hcrl_dataset}
\fontsize{8}{9.6}\selectfont
\setlength{\tabcolsep}{3pt}
\begin{tabular}{lccc}
\hline
Attack class & \begin{tabular}[c]{@{}c@{}}Number of\\normal messages\end{tabular} & \begin{tabular}[c]{@{}c@{}}Number of\\attack messages\end{tabular} & \begin{tabular}[c]{@{}c@{}}Total number\\of messages\end{tabular} \\
\hline
DoS & 3,078,250 & 587,521 & 3,665,771 \\
Fuzzy & 3,347,013 & 491,847 & 3,838,860 \\
Gear spoofing & 3,845,890 & 597,252 & 4,443,142 \\
RPM spoofing & 3,966,805 & 654,897 & 4,621,702 \\
\hline
\end{tabular}
\end{table}
\subsection{Experimental Setup}

The vehicle-side lightweight IDS models and proxy generators were evaluated on a workstation equipped with an NVIDIA GeForce RTX 4060 Laptop GPU and an Intel Core i7-13620H CPU. Raw CAN messages were preprocessed into windowed samples with $L=64$ using the method described in Section~\ref{sec:data_preprocessing}.

For each dataset, the raw CAN message stream was first divided into disjoint
blocks. These blocks were then assigned approximately equally to the training,
test, and online detection sets, after which windowed samples with $L=64$ were
constructed independently within each set. No window crossed a split boundary.
The training set was subsequently partitioned into five equal-sized local
training sets with class proportions drawn from a Dirichlet distribution with
$\alpha=0.5$.

Each vehicle used a bidirectional long short-term memory (BiLSTM) network with a serialized size below 1~MB as its lightweight IDS model. CANMentor can also use other lightweight multiclass models.

The confidence threshold and random high-confidence sampling ratio were set to 0.90 and 0.5\%, respectively. We set $(\lambda_{\mathrm{proxy}},\lambda_{\mathrm{clean}})=(1.0,0.3)$ for teacher adaptation and $(\tau,\lambda_{\mathrm{anchor}})=(2,0.2)$ for knowledge distillation.

\subsection{Cloud-Side Teacher Inference and Knowledge Distillation Results}
We evaluate Qwen2.5-1.5B, Qwen2.5-7B, and OPT-2.7B as cloud-side LLM teachers and compare the vehicle-side IDS models before and after knowledge distillation. We report accuracy and class-wise precision, recall, and F1, with macro averages summarizing class-level performance. Proxy samples use $d=0.40$ on the ROAD dataset and $d=0.35$ on the HCRL Car-Hacking dataset, as detailed in Section~\ref{sec:perturbation_utility_tradeoff}.

\subsubsection{ROAD Dataset Results}

Table~\ref{tab:road_original_client_results} reports the non-IID baseline results for all five vehicles, whose equal-sized training sets have different class proportions under the Dirichlet partition.
Throughout Tables~\ref{tab:road_original_client_results}--%
\ref{tab:hcrl_opt27b_classwise}, Acc denotes the overall accuracy for each vehicle. In the class-wise tables, P, R,
and F1 are reported per class, and the Macro row gives their unweighted
averages across classes.
\begin{table}[!t]
\centering
\caption{Original vehicle results on the ROAD dataset before distillation}
\fontsize{8}{9.6}\selectfont
\label{tab:road_original_client_results}
\begin{tabular}{lcccc}
\hline
Veh & Acc & Macro P & Macro R & Macro F1 \\
\hline
V0 & 0.9809 & 0.9814 & 0.9747 & 0.9775 \\
V1 & 0.9764 & 0.9743 & 0.9648 & 0.9687 \\
V2 & 0.9948 & 0.9955 & 0.9910 & 0.9932 \\
V3 & 0.9178 & 0.9614 & 0.8577 & 0.8399 \\
V4 & 0.6252 & 0.8144 & 0.8950 & 0.7778 \\
\hline
\end{tabular}
\end{table}

Vehicle 2 already achieves 0.9948 accuracy and 0.9932 Macro F1. In contrast, Vehicle 3 has only two Coolant training windows, while Vehicle 4 has few Normal and Fuzzing windows and only 0.6252 accuracy. We therefore evaluate distillation on Vehicles 0, 1, 3, and 4, whose class-specific blind spots vary in severity.

Tables~\ref{tab:road_qwen25_15b_classwise}--\ref{tab:road_opt27b_classwise}
report overall accuracy and class-wise precision, recall, and F1
on the test set before and after knowledge distillation on selected
windowed samples from the online detection set. N, Fuz, Cool, ROn, Corr, Spd, and ROff denote Normal, Fuzzing, Coolant, Reverse On, Correlated, Speedometer, and Reverse Off, respectively. The targeted attack classes use the masquerade setting. Changes within $\pm0.01$ remain black; larger improvements and decreases are shown in green and red, respectively, and (*) marks a large improvement with $\Delta\geq0.05$.


\begin{table}[!t]
\centering
\caption{Results on the ROAD dataset using Qwen2.5-1.5B as the cloud-side LLM teacher}
\label{tab:road_qwen25_15b_classwise}
\scriptsize
\setlength{\tabcolsep}{1pt}
\renewcommand{\arraystretch}{1.00}
\begin{tabular}{@{}llcccc@{}}
\hline
Veh & Class & \shortstack{Acc} & P & R & F1 \\
\hline
\multirow{8}{*}{V0} & N & \multirow{8}{*}{0.9809/\metricgain{0.9962}} & 0.9798/\metricgain{0.9993} & 0.9892/\metricsame{0.9961} & 0.9845/\metricgain{0.9977} \\
  & Fuz &  & 0.9895/\metricsame{0.9989} & 1.0000/\metricsame{1.0000} & 0.9947/\metricsame{0.9994} \\
  & Cool &  & 0.9996/\metricsame{1.0000} & 0.9826/\metricsame{0.9826} & 0.9910/\metricsame{0.9912} \\
  & ROn &  & 0.9554/\metricgain{0.9816} & 0.8514/\metricbiggain{0.9952} & 0.9004/\metricbiggain{0.9884} \\
  & Corr &  & 1.0000/\metricsame{1.0000} & 1.0000/\metricsame{1.0000} & 1.0000/\metricsame{1.0000} \\
  & Spd &  & 0.9797/\metricgain{0.9994} & 0.9995/\metricsame{0.9995} & 0.9895/\metricsame{0.9994} \\
  & ROff &  & 0.9658/\metricgain{0.9786} & 1.0000/\metricsame{1.0000} & 0.9826/\metricsame{0.9892} \\
  & Macro &  & 0.9814/\metricgain{0.9940} & 0.9747/\metricgain{0.9962} & 0.9775/\metricgain{0.9951} \\
\hline
\multirow{8}{*}{V1} & N & \multirow{8}{*}{0.9764/\metricgain{0.9938}} & 0.9800/\metricgain{0.9966} & 0.9919/\metricsame{0.9936} & 0.9859/\metricsame{0.9951} \\
  & Fuz &  & 0.9893/\metricsame{0.9949} & 0.8440/\metricbiggain{0.9750} & 0.9109/\metricbiggain{0.9848} \\
  & Cool &  & 0.9999/\metricsame{1.0000} & 1.0000/\metricsame{1.0000} & 0.9999/\metricsame{1.0000} \\
  & ROn &  & 0.9874/\metricsame{0.9960} & 0.9982/\metricsame{0.9999} & 0.9928/\metricsame{0.9979} \\
  & Corr &  & 1.0000/\metricsame{1.0000} & 1.0000/\metricsame{1.0000} & 1.0000/\metricsame{1.0000} \\
  & Spd &  & 0.9018/\metricbiggain{0.9690} & 0.9195/\metricbiggain{0.9892} & 0.9106/\metricbiggain{0.9790} \\
  & ROff &  & 0.9617/\metricgain{0.9866} & 1.0000/\metricsame{1.0000} & 0.9805/\metricgain{0.9932} \\
  & Macro &  & 0.9743/\metricgain{0.9919} & 0.9648/\metricgain{0.9940} & 0.9687/\metricgain{0.9929} \\
\hline
\multirow{8}{*}{V3} & N & \multirow{8}{*}{0.9178/\metricbiggain{0.9764}} & 0.8778/\metricbiggain{0.9613} & 0.9980/\metricsame{0.9977} & 0.9341/\metricgain{0.9792} \\
  & Fuz &  & 0.9607/\metricgain{1.0000} & 1.0000/\metricsame{1.0000} & 0.9800/\metricgain{1.0000} \\
  & Cool &  & 0.9804/\metricsame{0.9812} & 0.0063/\metricbiggain{0.7296} & 0.0124/\metricbiggain{0.8369} \\
  & ROn &  & 0.9988/\metricsame{0.9996} & 0.9995/\metricsame{0.9995} & 0.9991/\metricsame{0.9996} \\
  & Corr &  & 1.0000/\metricsame{0.9999} & 1.0000/\metricsame{1.0000} & 1.0000/\metricsame{0.9999} \\
  & Spd &  & 1.0000/\metricsame{1.0000} & 0.9999/\metricsame{0.9961} & 0.9999/\metricsame{0.9981} \\
  & ROff &  & 0.9122/\metricbiggain{0.9767} & 1.0000/\metricsame{1.0000} & 0.9541/\metricgain{0.9882} \\
  & Macro &  & 0.9614/\metricgain{0.9884} & 0.8577/\metricbiggain{0.9604} & 0.8399/\metricbiggain{0.9717} \\
\hline
\multirow{8}{*}{V4} & N & \multirow{8}{*}{0.6252/\metricbiggain{0.9703}} & 0.9998/\metricsame{0.9994} & 0.2655/\metricbiggain{0.9425} & 0.4196/\metricbiggain{0.9701} \\
  & Fuz &  & 0.9736/\metricgain{0.9994} & 1.0000/\metricsame{1.0000} & 0.9866/\metricgain{0.9997} \\
  & Cool &  & 0.9959/\metricsame{0.9995} & 1.0000/\metricsame{1.0000} & 0.9979/\metricsame{0.9998} \\
  & ROn &  & 0.2141/\metricbiggain{0.7790} & 0.9998/\metricsame{0.9965} & 0.3527/\metricbiggain{0.8744} \\
  & Corr &  & 0.9980/\metricsame{0.9996} & 1.0000/\metricsame{1.0000} & 0.9990/\metricsame{0.9998} \\
  & Spd &  & 0.9856/\metricgain{1.0000} & 0.9999/\metricsame{0.9990} & 0.9927/\metricsame{0.9995} \\
  & ROff &  & 0.5336/\metricbiggain{0.9291} & 1.0000/\metricsame{0.9998} & 0.6959/\metricbiggain{0.9632} \\
  & Macro &  & 0.8144/\metricbiggain{0.9580} & 0.8950/\metricbiggain{0.9911} & 0.7778/\metricbiggain{0.9723} \\
\hline
\end{tabular}
\end{table}

\begin{table}[!t]
\centering
\caption{Results on the ROAD dataset using Qwen2.5-7B as the cloud-side LLM teacher}
\label{tab:road_qwen25_7b_classwise}
\scriptsize
\setlength{\tabcolsep}{1pt}
\renewcommand{\arraystretch}{1.00}
\begin{tabular}{@{}llcccc@{}}
\hline
Veh & Class & \shortstack{Acc} & P & R & F1 \\
\hline
\multirow{8}{*}{V0} & N & \multirow{8}{*}{0.9809/\metricgain{0.9969}} & 0.9798/\metricgain{0.9996} & 0.9892/\metricsame{0.9973} & 0.9845/\metricgain{0.9984} \\
  & Fuz &  & 0.9895/\metricsame{0.9989} & 1.0000/\metricsame{1.0000} & 0.9947/\metricsame{0.9994} \\
  & Cool &  & 0.9996/\metricsame{1.0000} & 0.9826/\metricsame{0.9826} & 0.9910/\metricsame{0.9912} \\
  & ROn &  & 0.9554/\metricgain{0.9895} & 0.8514/\metricbiggain{0.9971} & 0.9004/\metricbiggain{0.9933} \\
  & Corr &  & 1.0000/\metricsame{1.0000} & 1.0000/\metricsame{1.0000} & 1.0000/\metricsame{1.0000} \\
  & Spd &  & 0.9797/\metricgain{0.9996} & 0.9995/\metricsame{0.9995} & 0.9895/\metricgain{0.9996} \\
  & ROff &  & 0.9658/\metricgain{0.9774} & 1.0000/\metricsame{1.0000} & 0.9826/\metricsame{0.9886} \\
  & Macro &  & 0.9814/\metricgain{0.9950} & 0.9747/\metricgain{0.9966} & 0.9775/\metricgain{0.9958} \\
\hline
\multirow{8}{*}{V1} & N & \multirow{8}{*}{0.9764/\metricgain{0.9953}} & 0.9800/\metricgain{0.9974} & 0.9919/\metricsame{0.9944} & 0.9859/\metricsame{0.9959} \\
  & Fuz &  & 0.9893/\metricsame{0.9924} & 0.8440/\metricbiggain{0.9852} & 0.9109/\metricbiggain{0.9888} \\
  & Cool &  & 0.9999/\metricsame{1.0000} & 1.0000/\metricsame{1.0000} & 0.9999/\metricsame{1.0000} \\
  & ROn &  & 0.9874/\metricgain{0.9975} & 0.9982/\metricsame{0.9990} & 0.9928/\metricsame{0.9983} \\
  & Corr &  & 1.0000/\metricsame{1.0000} & 1.0000/\metricsame{1.0000} & 1.0000/\metricsame{1.0000} \\
  & Spd &  & 0.9018/\metricbiggain{0.9733} & 0.9195/\metricbiggain{0.9931} & 0.9106/\metricbiggain{0.9831} \\
  & ROff &  & 0.9617/\metricgain{0.9959} & 1.0000/\metricsame{1.0000} & 0.9805/\metricgain{0.9979} \\
  & Macro &  & 0.9743/\metricgain{0.9938} & 0.9648/\metricgain{0.9960} & 0.9687/\metricgain{0.9949} \\
\hline
\multirow{8}{*}{V3} & N & \multirow{8}{*}{0.9178/\metricbiggain{0.9892}} & 0.8778/\metricbiggain{0.9847} & 0.9980/\metricsame{0.9952} & 0.9341/\metricbiggain{0.9899} \\
  & Fuz &  & 0.9607/\metricgain{0.9995} & 1.0000/\metricsame{1.0000} & 0.9800/\metricgain{0.9998} \\
  & Cool &  & 0.9804/\metricdrop{0.9699} & 0.0063/\metricbiggain{0.8986} & 0.0124/\metricbiggain{0.9329} \\
  & ROn &  & 0.9988/\metricsame{1.0000} & 0.9995/\metricsame{0.9992} & 0.9991/\metricsame{0.9996} \\
  & Corr &  & 1.0000/\metricsame{0.9985} & 1.0000/\metricsame{1.0000} & 1.0000/\metricsame{0.9993} \\
  & Spd &  & 1.0000/\metricsame{1.0000} & 0.9999/\metricsame{1.0000} & 0.9999/\metricsame{1.0000} \\
  & ROff &  & 0.9122/\metricbiggain{0.9940} & 1.0000/\metricsame{0.9996} & 0.9541/\metricgain{0.9968} \\
  & Macro &  & 0.9614/\metricgain{0.9924} & 0.8577/\metricbiggain{0.9847} & 0.8399/\metricbiggain{0.9883} \\
\hline
\multirow{8}{*}{V4} & N & \multirow{8}{*}{0.6252/\metricbiggain{0.9657}} & 0.9998/\metricsame{0.9993} & 0.2655/\metricbiggain{0.9336} & 0.4196/\metricbiggain{0.9654} \\
  & Fuz &  & 0.9736/\metricgain{0.9999} & 1.0000/\metricsame{1.0000} & 0.9866/\metricgain{0.9999} \\
  & Cool &  & 0.9959/\metricsame{1.0000} & 1.0000/\metricsame{1.0000} & 0.9979/\metricsame{1.0000} \\
  & ROn &  & 0.2141/\metricbiggain{0.7336} & 0.9998/\metricsame{0.9949} & 0.3527/\metricbiggain{0.8445} \\
  & Corr &  & 0.9980/\metricsame{0.9999} & 1.0000/\metricsame{1.0000} & 0.9990/\metricsame{0.9999} \\
  & Spd &  & 0.9856/\metricgain{0.9983} & 0.9999/\metricsame{1.0000} & 0.9927/\metricsame{0.9991} \\
  & ROff &  & 0.5336/\metricbiggain{0.9495} & 1.0000/\metricsame{0.9998} & 0.6959/\metricbiggain{0.9740} \\
  & Macro &  & 0.8144/\metricbiggain{0.9544} & 0.8950/\metricbiggain{0.9897} & 0.7778/\metricbiggain{0.9690} \\
\hline
\end{tabular}
\end{table}

\begin{table}[!t]
\centering
\caption{Results on the ROAD dataset using OPT-2.7B as the cloud-side LLM teacher}
\label{tab:road_opt27b_classwise}
\scriptsize
\setlength{\tabcolsep}{1pt}
\renewcommand{\arraystretch}{1.00}
\begin{tabular}{@{}llcccc@{}}
\hline
Veh & Class & \shortstack{Acc} & P & R & F1 \\
\hline
\multirow{8}{*}{V0} & N & \multirow{8}{*}{0.9809/\metricgain{0.9951}} & 0.9798/\metricgain{0.9993} & 0.9892/\metricsame{0.9943} & 0.9845/\metricgain{0.9968} \\
  & Fuz &  & 0.9895/\metricsame{0.9995} & 1.0000/\metricsame{1.0000} & 0.9947/\metricsame{0.9998} \\
  & Cool &  & 0.9996/\metricsame{1.0000} & 0.9826/\metricsame{0.9826} & 0.9910/\metricsame{0.9912} \\
  & ROn &  & 0.9554/\metricgain{0.9715} & 0.8514/\metricbiggain{0.9930} & 0.9004/\metricbiggain{0.9821} \\
  & Corr &  & 1.0000/\metricsame{1.0000} & 1.0000/\metricsame{1.0000} & 1.0000/\metricsame{1.0000} \\
  & Spd &  & 0.9797/\metricgain{0.9973} & 0.9995/\metricsame{0.9995} & 0.9895/\metricsame{0.9984} \\
  & ROff &  & 0.9658/\metricgain{0.9767} & 1.0000/\metricsame{1.0000} & 0.9826/\metricsame{0.9882} \\
  & Macro &  & 0.9814/\metricgain{0.9920} & 0.9747/\metricgain{0.9956} & 0.9775/\metricgain{0.9938} \\
\hline
\multirow{8}{*}{V1} & N & \multirow{8}{*}{0.9764/\metricgain{0.9933}} & 0.9800/\metricgain{0.9972} & 0.9919/\metricsame{0.9922} & 0.9859/\metricsame{0.9947} \\
  & Fuz &  & 0.9893/\metricsame{0.9939} & 0.8440/\metricbiggain{0.9766} & 0.9109/\metricbiggain{0.9852} \\
  & Cool &  & 0.9999/\metricsame{0.9999} & 1.0000/\metricsame{1.0000} & 0.9999/\metricsame{0.9999} \\
  & ROn &  & 0.9874/\metricsame{0.9933} & 0.9982/\metricsame{1.0000} & 0.9928/\metricsame{0.9966} \\
  & Corr &  & 1.0000/\metricsame{1.0000} & 1.0000/\metricsame{1.0000} & 1.0000/\metricsame{1.0000} \\
  & Spd &  & 0.9018/\metricbiggain{0.9636} & 0.9195/\metricbiggain{0.9908} & 0.9106/\metricbiggain{0.9770} \\
  & ROff &  & 0.9617/\metricgain{0.9864} & 1.0000/\metricsame{1.0000} & 0.9805/\metricgain{0.9932} \\
  & Macro &  & 0.9743/\metricgain{0.9906} & 0.9648/\metricgain{0.9942} & 0.9687/\metricgain{0.9924} \\
\hline
\multirow{8}{*}{V3} & N & \multirow{8}{*}{0.9178/\metricbiggain{0.9808}} & 0.8778/\metricbiggain{0.9694} & 0.9980/\metricsame{0.9974} & 0.9341/\metricgain{0.9832} \\
  & Fuz &  & 0.9607/\metricgain{0.9996} & 1.0000/\metricsame{1.0000} & 0.9800/\metricgain{0.9998} \\
  & Cool &  & 0.9804/\metricsame{0.9815} & 0.0063/\metricbiggain{0.7820} & 0.0124/\metricbiggain{0.8705} \\
  & ROn &  & 0.9988/\metricsame{1.0000} & 0.9995/\metricsame{0.9995} & 0.9991/\metricsame{0.9997} \\
  & Corr &  & 1.0000/\metricsame{0.9998} & 1.0000/\metricsame{1.0000} & 1.0000/\metricsame{0.9999} \\
  & Spd &  & 1.0000/\metricsame{1.0000} & 0.9999/\metricsame{0.9998} & 0.9999/\metricsame{0.9999} \\
  & ROff &  & 0.9122/\metricbiggain{0.9778} & 1.0000/\metricsame{1.0000} & 0.9541/\metricgain{0.9888} \\
  & Macro &  & 0.9614/\metricgain{0.9897} & 0.8577/\metricbiggain{0.9684} & 0.8399/\metricbiggain{0.9774} \\
\hline
\multirow{8}{*}{V4} & N & \multirow{8}{*}{0.6252/\metricbiggain{0.9663}} & 0.9998/\metricsame{0.9995} & 0.2655/\metricbiggain{0.9345} & 0.4196/\metricbiggain{0.9659} \\
  & Fuz &  & 0.9736/\metricgain{0.9994} & 1.0000/\metricsame{1.0000} & 0.9866/\metricgain{0.9997} \\
  & Cool &  & 0.9959/\metricsame{0.9903} & 1.0000/\metricsame{1.0000} & 0.9979/\metricsame{0.9951} \\
  & ROn &  & 0.2141/\metricbiggain{0.7501} & 0.9998/\metricsame{0.9970} & 0.3527/\metricbiggain{0.8561} \\
  & Corr &  & 0.9980/\metricsame{0.9996} & 1.0000/\metricsame{1.0000} & 0.9990/\metricsame{0.9998} \\
  & Spd &  & 0.9856/\metricgain{1.0000} & 0.9999/\metricsame{0.9995} & 0.9927/\metricsame{0.9997} \\
  & ROff &  & 0.5336/\metricbiggain{0.9370} & 1.0000/\metricsame{1.0000} & 0.6959/\metricbiggain{0.9675} \\
  & Macro &  & 0.8144/\metricbiggain{0.9537} & 0.8950/\metricbiggain{0.9901} & 0.7778/\metricbiggain{0.9691} \\
\hline
\end{tabular}
\end{table}

The largest ROAD gains occur for Vehicles 3 and 4. Vehicle 3's Coolant F1 rises from 0.0124 to 0.8369--0.9329 after distillation, reflecting repair of its severe blind spot. Its Macro F1 increases from 0.8399 to 0.9717--0.9883.

Vehicle 4's accuracy increases from 0.6252 to 0.9657--0.9703, while Macro F1 rises from 0.7778 to 0.9690--0.9723, with a maximum relative Macro F1 improvement of 25.01\%. Its Normal F1 improves from 0.4196 to 0.9654--0.9701, addressing a major initial weakness.

Vehicles 0 and 1 show smaller refinements, mainly in Reverse On for Vehicle 0 and Fuzzing and Speedometer for Vehicle 1; previously strong classes remain stable. The vehicle-side IDS models achieve their best results after
distillation from Qwen2.5-7B on Vehicles 0, 1, and 3,
and from Qwen2.5-1.5B on Vehicle 4. Thus, distillation gains do not depend solely on teacher size.

\subsubsection{HCRL Car-Hacking Dataset Results}

Table~\ref{tab:hcrl_original_client_results} reports the non-IID baseline results for all five vehicles.

\begin{table}[!t]
\centering
\caption{Original vehicle results on the HCRL Car-Hacking dataset before distillation}
\label{tab:hcrl_original_client_results}
\fontsize{8}{9.6}\selectfont
\begin{tabular}{lcccc}
\hline
Veh & Acc & Macro P & Macro R & Macro F1 \\
\hline
V0 & 0.9143 & 0.9647 & 0.8243 & 0.8258 \\
V1 & 0.9991 & 0.9996 & 0.9984 & 0.9990 \\
V2 & 0.9765 & 0.9573 & 0.9917 & 0.9732 \\
V3 & 0.9828 & 0.9831 & 0.9656 & 0.9728 \\
V4 & 0.9984 & 0.9977 & 0.9984 & 0.9980 \\
\hline
\end{tabular}
\end{table}

Vehicles 1 and 4 already achieve accuracies of 0.9991 and 0.9984, with Macro F1 close to 1.0, so we evaluate Vehicles 0, 2, and 3. Vehicle 0 has a pronounced DoS blind spot; Vehicle 2 has weaker DoS and Fuzzy precision than Vehicles 1 and 4, while scarce local Fuzzy samples limit Vehicle 3's recall.

Tables~\ref{tab:hcrl_qwen25_15b_classwise}--\ref{tab:hcrl_opt27b_classwise} use the same before/after and visual annotation conventions as the ROAD tables. N, DoS, Fuz, Gear, and RPM denote Normal, DoS, Fuzzy, Gear spoofing, and RPM spoofing, respectively.


\begin{table}[!t]
\centering
\caption{Results on the HCRL Car-Hacking dataset using Qwen2.5-1.5B as the cloud-side LLM teacher}
\label{tab:hcrl_qwen25_15b_classwise}
\scriptsize
\setlength{\tabcolsep}{1pt}
\renewcommand{\arraystretch}{1.00}
\begin{tabular}{@{}llcccc@{}}
\hline
Veh & Class & \shortstack{Acc} & P & R & F1 \\
\hline
\multirow{6}{*}{V0} & N & \multirow{6}{*}{0.9143/\metricbiggain{0.9971}} & 0.8848/\metricbiggain{0.9964} & 0.9999/\metricsame{0.9989} & 0.9388/\metricbiggain{0.9977} \\
  & DoS &  & 1.0000/\metricsame{1.0000} & 0.1260/\metricbiggain{0.9814} & 0.2238/\metricbiggain{0.9906} \\
  & Fuz &  & 0.9414/\metricbiggain{0.9992} & 0.9978/\metricsame{0.9978} & 0.9687/\metricgain{0.9985} \\
  & Gear &  & 0.9975/\metricsame{0.9935} & 0.9979/\metricsame{0.9982} & 0.9977/\metricsame{0.9959} \\
  & RPM &  & 0.9999/\metricsame{1.0000} & 1.0000/\metricsame{0.9995} & 0.9999/\metricsame{0.9997} \\
  & Macro &  & 0.9647/\metricgain{0.9978} & 0.8243/\metricbiggain{0.9952} & 0.8258/\metricbiggain{0.9965} \\
\hline
\multirow{6}{*}{V2} & N & \multirow{6}{*}{0.9765/\metricgain{0.9982}} & 0.9994/\metricsame{0.9989} & 0.9620/\metricgain{0.9982} & 0.9803/\metricgain{0.9985} \\
  & DoS &  & 0.8664/\metricbiggain{0.9937} & 0.9991/\metricsame{0.9982} & 0.9281/\metricbiggain{0.9959} \\
  & Fuz &  & 0.9373/\metricbiggain{0.9958} & 0.9989/\metricsame{0.9975} & 0.9671/\metricgain{0.9966} \\
  & Gear &  & 0.9846/\metricgain{0.9995} & 0.9984/\metricsame{0.9979} & 0.9914/\metricsame{0.9987} \\
  & RPM &  & 0.9986/\metricsame{1.0000} & 0.9999/\metricsame{0.9992} & 0.9993/\metricsame{0.9996} \\
  & Macro &  & 0.9573/\metricgain{0.9976} & 0.9917/\metricsame{0.9982} & 0.9732/\metricgain{0.9979} \\
\hline
\multirow{6}{*}{V3} & N & \multirow{6}{*}{0.9828/\metricgain{0.9982}} & 0.9831/\metricgain{0.9972} & 0.9991/\metricsame{0.9999} & 0.9911/\metricsame{0.9986} \\
  & DoS &  & 0.9741/\metricgain{1.0000} & 0.9985/\metricsame{0.9985} & 0.9862/\metricgain{0.9992} \\
  & Fuz &  & 1.0000/\metricsame{0.9999} & 0.8331/\metricbiggain{0.9870} & 0.9090/\metricbiggain{0.9934} \\
  & Gear &  & 0.9762/\metricgain{1.0000} & 0.9991/\metricsame{0.9984} & 0.9875/\metricgain{0.9992} \\
  & RPM &  & 0.9822/\metricgain{0.9997} & 0.9982/\metricsame{0.9985} & 0.9901/\metricsame{0.9991} \\
  & Macro &  & 0.9831/\metricgain{0.9994} & 0.9656/\metricgain{0.9965} & 0.9728/\metricgain{0.9979} \\
\hline
\end{tabular}
\end{table}

\begin{table}[!t]
\centering
\caption{Results on the HCRL Car-Hacking dataset using Qwen2.5-7B as the cloud-side LLM teacher}
\label{tab:hcrl_qwen25_7b_classwise}
\scriptsize
\setlength{\tabcolsep}{1pt}
\renewcommand{\arraystretch}{1.00}
\begin{tabular}{@{}llcccc@{}}
\hline
Veh & Class & \shortstack{Acc} & P & R & F1 \\
\hline
\multirow{6}{*}{V0} & N & \multirow{6}{*}{0.9143/\metricbiggain{0.9980}} & 0.8848/\metricbiggain{0.9971} & 0.9999/\metricsame{0.9996} & 0.9388/\metricbiggain{0.9984} \\
  & DoS &  & 1.0000/\metricsame{1.0000} & 0.1260/\metricbiggain{0.9871} & 0.2238/\metricbiggain{0.9935} \\
  & Fuz &  & 0.9414/\metricbiggain{0.9976} & 0.9978/\metricsame{0.9979} & 0.9687/\metricgain{0.9978} \\
  & Gear &  & 0.9975/\metricsame{0.9997} & 0.9979/\metricsame{0.9968} & 0.9977/\metricsame{0.9982} \\
  & RPM &  & 0.9999/\metricsame{1.0000} & 1.0000/\metricsame{0.9996} & 0.9999/\metricsame{0.9998} \\
  & Macro &  & 0.9647/\metricgain{0.9989} & 0.8243/\metricbiggain{0.9962} & 0.8258/\metricbiggain{0.9975} \\
\hline
\multirow{6}{*}{V2} & N & \multirow{6}{*}{0.9765/\metricgain{0.9983}} & 0.9994/\metricsame{0.9988} & 0.9620/\metricgain{0.9985} & 0.9803/\metricgain{0.9986} \\
  & DoS &  & 0.8664/\metricbiggain{0.9963} & 0.9991/\metricsame{0.9981} & 0.9281/\metricbiggain{0.9972} \\
  & Fuz &  & 0.9373/\metricbiggain{0.9944} & 0.9989/\metricsame{0.9976} & 0.9671/\metricgain{0.9960} \\
  & Gear &  & 0.9846/\metricgain{0.9999} & 0.9984/\metricsame{0.9974} & 0.9914/\metricsame{0.9986} \\
  & RPM &  & 0.9986/\metricsame{1.0000} & 0.9999/\metricsame{0.9994} & 0.9993/\metricsame{0.9997} \\
  & Macro &  & 0.9573/\metricgain{0.9979} & 0.9917/\metricsame{0.9982} & 0.9732/\metricgain{0.9980} \\
\hline
\multirow{6}{*}{V3} & N & \multirow{6}{*}{0.9828/\metricgain{0.9984}} & 0.9831/\metricgain{0.9975} & 0.9991/\metricsame{0.9999} & 0.9911/\metricsame{0.9987} \\
  & DoS &  & 0.9741/\metricgain{1.0000} & 0.9985/\metricsame{0.9984} & 0.9862/\metricgain{0.9992} \\
  & Fuz &  & 1.0000/\metricsame{0.9991} & 0.8331/\metricbiggain{0.9890} & 0.9090/\metricbiggain{0.9940} \\
  & Gear &  & 0.9762/\metricgain{1.0000} & 0.9991/\metricsame{0.9984} & 0.9875/\metricgain{0.9992} \\
  & RPM &  & 0.9822/\metricgain{1.0000} & 0.9982/\metricsame{0.9981} & 0.9901/\metricsame{0.9991} \\
  & Macro &  & 0.9831/\metricgain{0.9993} & 0.9656/\metricgain{0.9968} & 0.9728/\metricgain{0.9980} \\
\hline
\end{tabular}
\end{table}

\begin{table}[!t]
\centering
\caption{Results on the HCRL Car-Hacking dataset using OPT-2.7B as the cloud-side LLM teacher}
\label{tab:hcrl_opt27b_classwise}
\scriptsize
\setlength{\tabcolsep}{1pt}
\renewcommand{\arraystretch}{1.00}
\begin{tabular}{@{}llcccc@{}}
\hline
Veh & Class & \shortstack{Acc} & P & R & F1 \\
\hline
\multirow{6}{*}{V0} & N & \multirow{6}{*}{0.9143/\metricbiggain{0.9970}} & 0.8848/\metricbiggain{0.9969} & 0.9999/\metricsame{0.9982} & 0.9388/\metricbiggain{0.9976} \\
  & DoS &  & 1.0000/\metricsame{1.0000} & 0.1260/\metricbiggain{0.9850} & 0.2238/\metricbiggain{0.9924} \\
  & Fuz &  & 0.9414/\metricbiggain{0.9997} & 0.9978/\metricsame{0.9971} & 0.9687/\metricgain{0.9984} \\
  & Gear &  & 0.9975/\metricsame{0.9887} & 0.9979/\metricsame{0.9986} & 0.9977/\metricsame{0.9937} \\
  & RPM &  & 0.9999/\metricsame{1.0000} & 1.0000/\metricsame{0.9995} & 0.9999/\metricsame{0.9997} \\
  & Macro &  & 0.9647/\metricgain{0.9971} & 0.8243/\metricbiggain{0.9957} & 0.8258/\metricbiggain{0.9964} \\
\hline
\multirow{6}{*}{V2} & N & \multirow{6}{*}{0.9765/\metricgain{0.9982}} & 0.9994/\metricsame{0.9987} & 0.9620/\metricgain{0.9984} & 0.9803/\metricgain{0.9986} \\
  & DoS &  & 0.8664/\metricbiggain{0.9963} & 0.9991/\metricsame{0.9982} & 0.9281/\metricbiggain{0.9973} \\
  & Fuz &  & 0.9373/\metricbiggain{0.9940} & 0.9989/\metricsame{0.9979} & 0.9671/\metricgain{0.9959} \\
  & Gear &  & 0.9846/\metricgain{0.9999} & 0.9984/\metricsame{0.9965} & 0.9914/\metricsame{0.9982} \\
  & RPM &  & 0.9986/\metricsame{1.0000} & 0.9999/\metricsame{0.9992} & 0.9993/\metricsame{0.9996} \\
  & Macro &  & 0.9573/\metricgain{0.9978} & 0.9917/\metricsame{0.9981} & 0.9732/\metricgain{0.9979} \\
\hline
\multirow{6}{*}{V3} & N & \multirow{6}{*}{0.9828/\metricgain{0.9983}} & 0.9831/\metricgain{0.9973} & 0.9991/\metricsame{0.9999} & 0.9911/\metricsame{0.9986} \\
  & DoS &  & 0.9741/\metricgain{1.0000} & 0.9985/\metricsame{0.9981} & 0.9862/\metricgain{0.9991} \\
  & Fuz &  & 1.0000/\metricsame{0.9990} & 0.8331/\metricbiggain{0.9885} & 0.9090/\metricbiggain{0.9937} \\
  & Gear &  & 0.9762/\metricgain{1.0000} & 0.9991/\metricsame{0.9984} & 0.9875/\metricgain{0.9992} \\
  & RPM &  & 0.9822/\metricgain{1.0000} & 0.9982/\metricsame{0.9981} & 0.9901/\metricsame{0.9991} \\
  & Macro &  & 0.9831/\metricgain{0.9993} & 0.9656/\metricgain{0.9966} & 0.9728/\metricgain{0.9979} \\
\hline
\end{tabular}
\end{table}

Vehicle 0 benefits most: all three teachers raise its DoS recall from 0.1260 to 0.9814--0.9871 and F1 from 0.2238 to 0.9906--0.9935. Its Macro F1 increases from 0.8258 to 0.9964--0.9975, with a maximum relative improvement of 20.79\%.

Vehicles 2 and 3 also improve: Vehicle 2's Macro F1 rises from 0.9732 to 0.9979--0.9980, and Vehicle 3's Fuzzy F1 rises from 0.9090 to 0.9934--0.9940. All three teachers improve the evaluated vehicles. The vehicle-side IDS models achieve the highest accuracy and
Macro F1 after distillation from Qwen2.5-7B in each case,
although the differences among distillation settings are small.
\subsubsection{Teacher Performance on Selected Windowed Samples}

We compare local predictions on selected windowed samples
with teacher predictions on the corresponding proxy samples. Randomly selected high-confidence samples are excluded, and labels are used only for offline evaluation. Table~\ref{tab:selected_sample_teacher_results} reports the results.

\begingroup
\setlength{\intextsep}{6pt}
\begin{table}[H]
\centering
\caption{Teacher performance on selected windowed samples}
\label{tab:selected_sample_teacher_results}
\fontsize{7}{8.4}\selectfont
\setlength{\tabcolsep}{2pt}
\renewcommand{\arraystretch}{1.08}
\begin{tabular}{llrccr}
\toprule
Dataset & Teacher & $N$
& \begin{tabular}[c]{@{}c@{}}Local Acc\\(\%)\end{tabular}
& \begin{tabular}[c]{@{}c@{}}Teacher Acc\\(\%)\end{tabular}
& Gain (pp) \\
\midrule
\multirow{3}{*}{ROAD}
& Qwen2.5-1.5B & 6,192 & 55.91 & 96.95 & +41.04 \\
& OPT-2.7B     & 6,155 & 57.55 & 96.26 & +38.71 \\
& Qwen2.5-7B   & 5,840 & 56.01 & 95.70 & +39.69 \\
\midrule
\multirow{3}{*}{HCRL Car-Hacking}
& Qwen2.5-1.5B & 3,846 & 54.60 & 96.13 & +41.53 \\
& OPT-2.7B     & 3,828 & 54.81 & 96.81 & +42.00 \\
& Qwen2.5-7B   & 3,947 & 55.13 & 96.55 & +41.42 \\
\bottomrule
\end{tabular}
\par\vspace{1mm}
\begin{minipage}{\columnwidth}
\textit{Note:} $N$ is the number of low-confidence windowed samples selected
from the online detection set. Local Acc is computed on the selected windowed
samples retained on the vehicle, whereas Teacher Acc is computed on their
corresponding proxy samples.
\end{minipage}
\end{table}
\endgroup

Local accuracy is relatively low because these windowed samples lie near the local decision boundary. Despite their difficulty and the perturbations introduced into the corresponding proxy samples, the three teachers achieve 95.70\%--96.95\% accuracy and outperform the local predictions by 38.71--42.00 percentage points across the two datasets. These results support using a timely teacher prediction as the current detection result while retaining all returned teacher logits for periodic knowledge distillation.
\begin{figure*}[!t]
\centering
\subfloat[]{
\includegraphics[width=0.485\textwidth]
{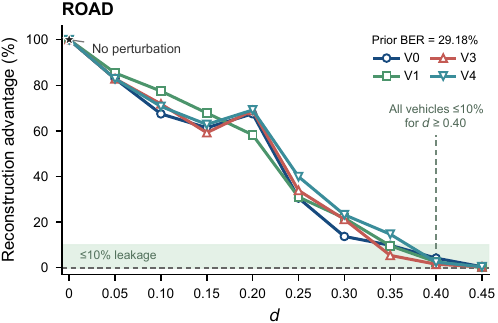}
\label{fig:road_reconstruction_leakage}}
\hfill
\subfloat[]{
\includegraphics[width=0.485\textwidth]
{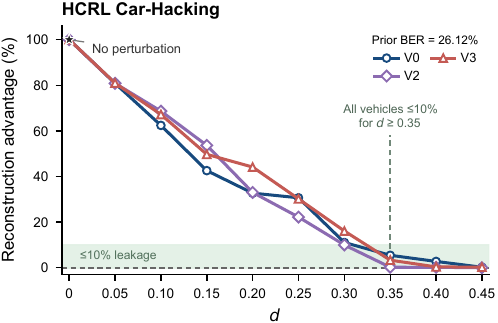}
\label{fig:hcrl_reconstruction_leakage}}

\vspace{-2mm}

\subfloat[]{
\includegraphics[width=0.485\textwidth]
{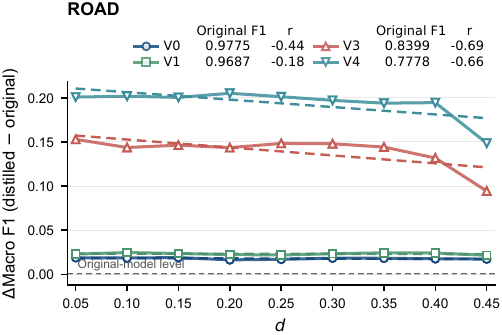}
\label{fig:road_distillation_gain}}
\hfill
\subfloat[]{
\includegraphics[width=0.485\textwidth]
{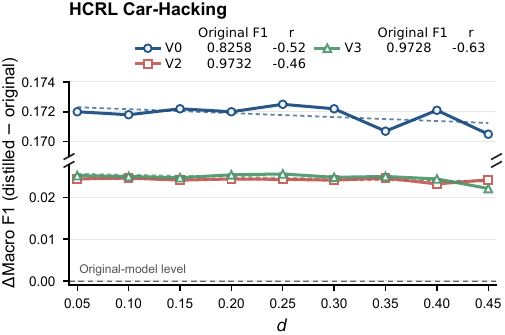}
\label{fig:hcrl_distillation_gain}}

\caption{Effect of the fixed perturbation distance on (a, b) reconstruction advantage and (c, d) change in vehicle-side Macro F1 after distillation from Qwen2.5-1.5B. Panels (a, c) show ROAD results; panels (b, d) show HCRL Car-Hacking results. Dashed lines show linear trends, and $r$ denotes the Pearson correlation coefficient.}

\label{fig:perturbation_utility}
\end{figure*}
\subsection{Perturbation--Privacy and Utility Trade-off Analysis}
\label{sec:perturbation_utility_tradeoff}
We evaluate two external reconstruction attacks: perturbation-matched reconstruction and architecture-aware shadow-generator transfer. Both use public CAN data and uploaded proxy samples, without query access to the deployed generators.

\subsubsection{Perturbation-Matched Reconstruction Attack}
To evaluate the privacy of uploaded proxy samples, we consider a reconstruction
adversary that has access to a separate public dataset from the same application
domain and the proxy samples uploaded by the vehicles. Following denoising-based reconstruction approaches~\cite{zhang2017beyond,wang2025task},
the adversary trains reconstruction models using synthetically perturbed public data at perturbation
levels matched to those of the observed proxy samples. However, it cannot access the target
vehicle's original CAN messages or the parameters of the proxy generator. This training encourages the models to learn regularities in the underlying CAN data to support the reconstruction of original samples from perturbed inputs.
Reconstruction performance is measured using the bit error rate (BER).

We evaluate a temporal convolutional neural network (T-CNN), BiLSTM, Transformer, and CNN--BiLSTM. For each vehicle and distance, the lowest-BER reconstructor defines the strongest learned attack, evaluated against the no-proxy per-bit prior.

Fig.~\ref{fig:perturbation_utility} shows reconstruction advantage and distillation performance as functions of $d$. We define reconstruction advantage as the percentage reduction in BER achieved by the best of the four learned reconstructors relative to the no-proxy per-bit prior, setting negative reductions to zero. This measure represents reconstruction leakage relative to the
no-proxy prior. Panels (a) and (b) report this measure for each vehicle.
\begin{table*}[!t]
\centering
\caption{Reconstruction attack results at $d=0.40$ on the ROAD dataset and $d=0.35$ on the HCRL Car-Hacking dataset. Lower BER indicates a stronger reconstruction attack.}
\label{tab:a1_reconstruction_audit}
\fontsize{8}{9.6}\selectfont
\setlength{\tabcolsep}{3.5pt}
\renewcommand{\arraystretch}{1.10}
\begin{tabular}{llcccccc}
\toprule
Dataset & Veh & Prior BER & T-CNN BER & BiLSTM BER & Transformer BER & CNN--BiLSTM BER & \shortstack{Reconstruction\\advantage (\%)} \\
\midrule
\multirow{4}{*}{ROAD}
& V0 & 0.2918 & 0.2878 & 0.2904 & \textbf{0.2795} & 0.2901 & 4.22 \\
& V1 & 0.2918 & 0.2884 & 0.2902 & \textbf{0.2843} & 0.2907 & 2.57 \\
& V3 & 0.2918 & \textbf{0.2873} & 0.2885 & 0.3124 & 0.2897 & 1.54 \\
& V4 & 0.2918 & \textbf{0.2847} & 0.2883 & 0.2912 & 0.2883 & 2.43 \\
\midrule
\multirow{3}{*}{HCRL Car-Hacking}
& V0 & 0.2612 & 0.2517 & 0.2597 & \textbf{0.2469} & 0.2598 & 5.47 \\
& V2 & 0.2612 & \textbf{0.2606} & 0.2614 & 0.2614 & 0.2616 & 0.23 \\
& V3 & 0.2612 & 0.2798 & 0.2630 & 0.2587 & \textbf{0.2525} & 3.33 \\
\bottomrule
\end{tabular}
\vspace{1mm}

\begin{minipage}{0.98\textwidth}
\footnotesize
\textit{Note:} Bold values indicate the lowest BER among the four learned reconstructors. Reconstruction advantage is calculated as $100\%\times(\mathrm{BER}_{\mathrm{prior}}-\mathrm{BER}_{\mathrm{strongest}})/\mathrm{BER}_{\mathrm{prior}}$, where $\mathrm{BER}_{\mathrm{prior}}$ is the no-proxy prior BER and $\mathrm{BER}_{\mathrm{strongest}}$ is the lowest BER among the four learned reconstructors. Veh. denotes vehicle (V0 = Vehicle 0, etc.).
\end{minipage}
\end{table*}

Reconstruction advantage is at most 4.22\% for ROAD ($d=0.40$) and 5.47\% for
HCRL Car-Hacking ($d=0.35$), and at most 0.43\% and 0.26\%,
respectively, at $d=0.45$. All evaluated vehicles remain below 10\% leakage at and above these distances within the tested range. Overall, reconstruction advantage decreases as $d$ increases.

Panels (c) and (d) show that Qwen2.5-1.5B distillation gains generally decrease with $d$, with vehicle-dependent sensitivity.

We select $d=0.40$ for ROAD and $d=0.35$ for HCRL Car-Hacking.
Both settings already yield low reconstruction advantage, while increasing
$d$ to 0.45 further reduces reconstruction advantage but lowers the mean Macro F1 gain.

Table~\ref{tab:a1_reconstruction_audit} details the four learned reconstructors at the selected distances. Their lowest BERs range from 0.2795 to 0.2873 on the ROAD dataset and from 0.2469 to 0.2606 on the HCRL Car-Hacking dataset, against prior BERs of 0.2918 and 0.2612, respectively.

\subsubsection{Architecture-Aware Shadow-Generator Transfer Attack}

We further consider an architecture-aware adversary that knows the generator
architecture, training procedure, and perturbation levels. Using only public CAN data, the adversary trains a surrogate IDS
and a shadow generator, constructs public proxy--original pairs,
and trains the same four reconstructors. The learned reconstructors are then directly applied to the proxy samples uploaded by the target vehicles, without access to the
target generators' parameters.

The lowest BERs range from 0.3283 to 0.3740 on ROAD
and from 0.3174 to 0.3518 on HCRL Car-Hacking.
Both ranges exceed the respective no-proxy prior BERs
of 0.2918 and 0.2612.

To further examine these results, we compare BERs on
HCRL Car-Hacking. The BiLSTM reconstructor achieves a BER of 0.0915 on validation
pairs constructed using the shadow generator, but 0.4179 on
target proxy samples from Vehicle V0. This indicates that low reconstruction error in the shadow domain does not translate into a reconstruction advantage in the target domain. Differences in the mappings and data distributions between the two domains may limit the transfer of reconstruction capabilities. Under our deployment assumptions, vehicles retain their original CAN data locally, and the external adversary cannot access proxy--original pairs from the target generator. It therefore cannot use such paired supervision to directly learn a reconstruction mapping specific to the target generator.
\subsection{Ablation Study}

We conduct two ablation studies to examine the contribution of the cloud-side
teacher design. First, we replace the LLM teacher with BiLSTM teachers of
different sizes to determine whether increasing the capacity of a conventional
teacher is sufficient to reproduce the distillation benefit. Second, we remove
teacher adaptation to proxy samples while retaining the numerical input
representation.
\subsubsection{Effect of Cloud-Side Teacher Selection}

We replace the cloud-side LLM teacher with BiLSTM teachers based on the
vehicle-side IDS architecture. Three configurations with serialized sizes of
approximately 100~MB, 1~GB, and 3~GB are evaluated alongside Qwen2.5-1.5B, the
smallest LLM teacher in the main experiments, with a serialized size of
approximately 3.1~GB. The 3~GB BiLSTM provides a size-comparable reference for
determining whether conventional teacher capacity alone can reproduce the
benefit of the LLM teacher.

\begin{figure}[!t]
\centering
\includegraphics[width=\columnwidth]{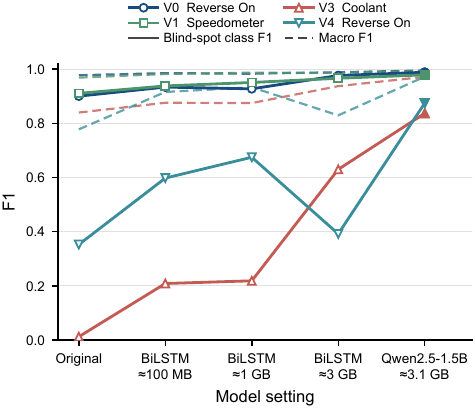}
\caption{Effect of teacher selection on the ROAD dataset. Solid lines show the
F1 of each vehicle's blind-spot class, while dashed lines show
the corresponding Macro F1. Results for each teacher represent
vehicle-side IDS performance after distillation; Original denotes
the pre-distillation baseline. The blind-spot classes are identified
before distillation: Reverse On for Vehicles 0 and 4, Speedometer
for Vehicle 1, and Coolant for Vehicle 3.}
\label{fig:teacher_selection_ablation}
\end{figure}

\begin{figure}[!t]
\centering
\includegraphics[
    width=0.88\columnwidth,
    height=0.30\textheight,
    keepaspectratio
]{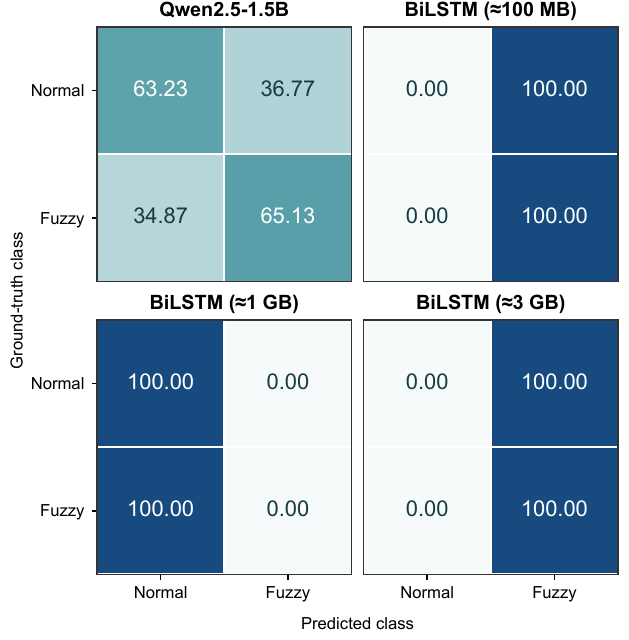}
\caption{HCRL-to-ROAD Normal/Fuzzy transfer without ROAD windowed samples for
training. The Fuzzy label denotes the HCRL Fuzzy class and the ROAD Fuzzing class. Each confusion matrix is normalized by ground-truth class, and entries are percentages.}
\label{fig:cross_dataset_generalization}
\end{figure}

In Fig.~\ref{fig:teacher_selection_ablation}, each vehicle-side
IDS model achieves its highest blind-spot F1 and Macro F1
after distillation from Qwen2.5-1.5B. Blind-spot F1 on Vehicles
3 and 4 rises from 0.012 and 0.353 to 0.837 and 0.874, respectively. Student Macro F1 exceeds 0.97 on both vehicles, compared with 0.937 and 0.829 after distillation from the 3~GB BiLSTM. Increasing BiLSTM size does not produce monotonic gains across vehicles; the difference between the two teachers of similar size indicates that size alone does not explain the student results.

To assess robustness beyond in-domain ROAD performance, we conduct an HCRL-to-ROAD stress test. We map the HCRL Fuzzy class to the ROAD Fuzzing class and use Fuzzy as their shared label. Because Normal and Fuzzy are the only classes shared by both datasets, the teachers are trained only on the corresponding windowed samples from the HCRL Car-Hacking dataset and evaluated directly on the ROAD dataset without using ROAD windowed samples for training. Fig.~\ref{fig:cross_dataset_generalization} compares Qwen2.5-1.5B with the 100~MB, 1~GB, and 3~GB BiLSTM teachers.

All three BiLSTM teachers collapse to a single-class prediction under the
cross-dataset shift. The 100~MB and 3~GB models classify every ROAD windowed
sample as Fuzzy, whereas the 1~GB model classifies every sample as Normal. Each
therefore obtains 50\% accuracy on the balanced test set and completely misses
one class. In contrast, Qwen2.5-1.5B retains discrimination between the
classes, achieving recalls of 63.23\% and 65.13\% for Normal and Fuzzy,
respectively, and an overall accuracy of 64.18\%. This size-matched comparison further supports the cross-dataset
robustness advantage of the LLM teacher.

\subsubsection{Effect of Teacher Adaptation to Proxy Samples}
\begin{figure*}[!t]
\centering
\includegraphics[width=\textwidth]{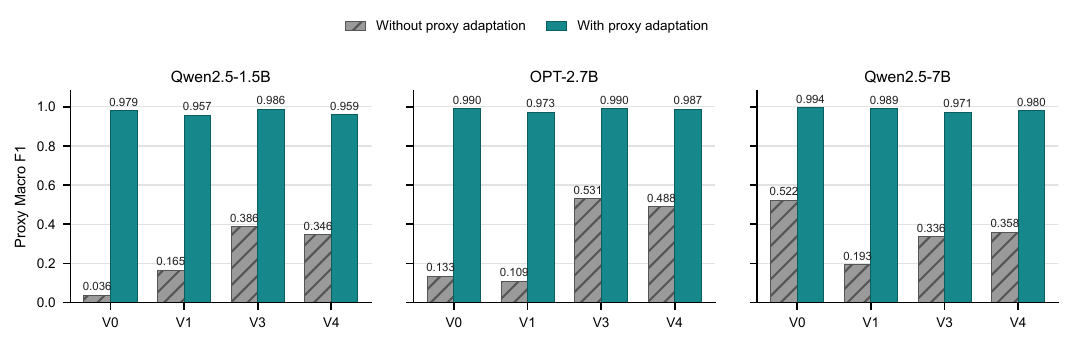}
\caption{Macro F1 on proxy samples with and without teacher adaptation
on ROAD ($d=0.40$) for Vehicles 0, 1, 3, and 4.
All settings use the field-aware numerical input representation.
Teachers from left to right are Qwen2.5-1.5B, OPT-2.7B, and Qwen2.5-7B.}
\label{fig:proxy_adaptation_ablation}
\end{figure*}

We next evaluate whether the cloud-side LLM teachers can perform reliable
inference on proxy samples without additional adaptation. For each teacher, we compare the settings with and without proxy adaptation
using the same generators and evaluation samples (Fig.~\ref{fig:proxy_adaptation_ablation}).

Without adaptation to proxy samples, the mean Macro F1 values are only 0.233, 0.315, and 0.352 for Qwen2.5-1.5B, OPT-2.7B, and Qwen2.5-7B, respectively. After adaptation, the corresponding mean values increase to 0.970, 0.985,
and 0.984, with all per-vehicle Macro F1 values reaching at least 0.957. Although the numerical encoder addresses the mismatch between CAN signal fields and the LLM input space, proxy adaptation remains necessary to accommodate generator perturbations. The high Macro F1 after adaptation indicates that proxy samples
retain information useful for distinguishing normal traffic from
attacks and differentiating attack classes. Adaptation enables
the teacher to use this information despite the distribution
shift introduced by the proxy generator.

\subsection{Runtime Overhead Analysis}
\begin{table}[!t]
\centering
\caption{Runtime overhead of CANMentor on the ROAD dataset and the HCRL Car-Hacking dataset (in milliseconds)}
\label{tab:runtime_overhead}
\small
\setlength{\tabcolsep}{6pt}
\renewcommand{\arraystretch}{1.12}
\begin{tabular}{lcc}
\hline
Operation & ROAD & HCRL Car-Hacking \\
\hline
\multicolumn{3}{l}{\textit{Vehicle-side processing time}} \\
IDS inference & 0.60 & 0.59 \\
Proxy sample generation & 3.63 & 2.85 \\
\midrule
\multicolumn{3}{l}{\textit{Teacher inference time}} \\
Qwen2.5-1.5B & 27.04 & 28.51 \\
OPT-2.7B & 27.53 & 28.62 \\
Qwen2.5-7B & 30.85 & 34.22 \\
\hline

\end{tabular}
\end{table}

Table~\ref{tab:runtime_overhead} reports mean latencies per windowed
sample, excluding communication. IDS inference takes 0.60 and 0.59~ms
on ROAD and HCRL Car-Hacking, respectively, while proxy generation
takes 3.63 and 2.85~ms. Proxy generation is performed only for selected windowed samples
and is not required to obtain the local prediction.
Cloud-side teacher inference, including the numerical encoder, LLM,
and classification head, is reported separately. Actual response time also includes the round-trip communication
time $t_{\mathrm{comm},i}^{(u)}$, which depends on network conditions.
If the total response time exceeds $T_{\max}$, CANMentor retains
the local prediction and uses returned logits for subsequent
distillation.

\section{Discussion}
Results show that cloud-side LLM teachers improve vehicle-side IDS models under non-IID data through privacy-preserving proxies that keep original CAN messages local. For online detection, CANMentor uses teacher predictions only
when responses arrive within the latency bound; otherwise,
it retains the local prediction. Returned logits also support periodic distillation, enabling cloud assistance without continuous dependence on cloud availability.

Real-vehicle and larger-scale IoV evaluations can examine communication delay, intermittent cloud availability, and increased vehicle participation. Adaptive perturbation control and teacher selection can account for variations in vehicle data distributions, network conditions, and teacher models. Broader evaluations across vehicle platforms, attack classes, CAN datasets, and lightweight IDS architectures can further assess generalizability. The cloud-side LLM teacher can also be explored for attack explanation and cross-vehicle alert analysis.

\section{Conclusion}

CANMentor combines vehicle-side lightweight IDS models with a cloud-side LLM teacher through fixed-distance proxy samples. Teacher predictions are used only within the prescribed latency bound; otherwise, local predictions are retained. All returned teacher logits are converted into soft targets for periodic knowledge distillation.

Experiments on the ROAD dataset and the HCRL Car-Hacking dataset show that CANMentor improves vehicle-side IDS models with blind spots for underrepresented attack classes under non-IID data. The maximum relative Macro F1 improvements reach 25.01\% and 20.79\%, respectively. At $d=0.40$ on the ROAD dataset and $d=0.35$ on the HCRL Car-Hacking dataset,
the maximum reconstruction advantage is 4.22\% and 5.47\%,
respectively. The cloud-side teachers also achieve 95.70\%--96.95\% accuracy on proxy samples corresponding to selected
low-confidence windows.

\bibliographystyle{IEEEtranDOI}
\bibliography{references_ieee_clean}

\newpage

\vfill

\end{document}